%% file: stereoCP-v5.0.tex
\documentclass[useAMS,usenatbib]{mn2e}
\usepackage{txfonts}
\usepackage{longtable}  
\usepackage{rotating}
\usepackage{natbib}
\usepackage{graphicx}
\usepackage{graphics}
\usepackage{psfrag}
\usepackage{amssymb}
\usepackage{multicol}
\usepackage{lscape}
\bibpunct{(}{)}{;}{a}{}{,}
\def\Teff{\ensuremath{T_{\mathrm{eff}}}}
\def\logg{\ensuremath{\log g}}

\def\vsini{\ensuremath{{\upsilon}\sin i}}
\def\kms{$\mathrm{km\,s}^{-1}$}

\def\veq{$V_{\mathrm{eq}}$}

\def\espa{ESPaDOnS}

\def\llm{{\sc LLmodels}}

\def\logl{\ensuremath{\log L/L_{\odot}}}

\def\M{\ensuremath{M/M_{\odot}}}

\def\stereo{\textit {STEREO}}
\def\asec{\hbox{"\hskip-3pt\,\,}}
\title[A \stereo\ photometric study of chemically peculiar stars]
{A photometric study of chemically peculiar stars with the \textit{STEREO} 
satellites. I. Magnetic chemically peculiar stars\thanks{Data obtained with 
the Heliospheric Imager instruments on board the \stereo\ spacecraft.}} 

\author[K.\,T. Wraight et al.]{K.\,T. Wraight,$^{1}$\thanks{K.T.Wraight@open.ac.uk}
			   L. Fossati,$^{1}$
			   M. Netopil,$^{2}$
			   E. Paunzen,$^{2}$
			   M. Rode-Paunzen,$^{2}$
	       \newauthor  D. Bewsher,$^{3}$ 
	       		   A.\,J. Norton$^{1}$ and
			   G.\,J. White$^{1,4}$\\
$^{1}$Department of Physical Sciences, Open University, 
Walton Hall, Milton Keynes MK7 6AA, UK\\
$^{2}$Institut f\"ur Astronomie, Universit\"{a}t Wien, 
T\"{u}rkenschanzstrasse 17, 1180 Wien, Austria\\
$^{3}$Jeremiah Horrocks Institute, University of Central 
Lancashire, Preston, Lancashire, PR1 2HE, UK\\
$^{4}$Space Science and Technology Department, STFC Rutherford 
Appleton Laboratory, Chilton, Didcot, Oxfordshire, OX11 0QX, UK}
\begin{document}

\date{}

\pagerange{\pageref{firstpage}--\pageref{lastpage}} \pubyear{2011}

\maketitle

\label{firstpage}

\begin{abstract}
About 10\,\% of upper main sequence stars are characterised by the presence of
chemical peculiarities, often found together with a structured magnetic field. 
The atmospheres of most of those chemically peculiar stars present surface 
spots, leading to photometric variability caused by rotational modulation. The 
study of the light curves of those stars therefore, permits a precise 
measurement of their rotational period, which is important to study stellar 
evolution and to plan further detailed observations. We analysed the light 
curves of 1028 chemically peculiar stars obtained with the \stereo\ spacecraft. 
We present here the results obtained for the 337 magnetic chemically peculiar 
stars in our sample. Thanks to the cadence and stability of the photometry, 
\stereo\ data are perfectly suitable to study variability signals with a 
periodicity typical of magnetic chemically peculiar stars. Using a 
matched filter algorithm and then two different period searching algorithms, 
we compiled a list of 82 magnetic chemically peculiar stars for which we 
measured a reliable rotational period; for 48 of them this is the first
measurement of their rotational period. The remaining 255 stars are likely to 
be constant, although we cannot exclude the presence of long period 
variability. In some cases, the presence of blending or systematic effects 
prevented us from detecting any reliable variability and in those cases we 
classified the star as constant. For each star we classified as variable, we 
determined temperature, luminosity, mass and fractional age, but the limited 
statistics, biased towards the shorter periods, prevented us from finding 
any evolutionary trend of the rotational period. For a few stars, the 
comparison between their projected rotational velocity and equatorial velocity 
let us believe that their real rotational period might be longer than that 
found here and previously obtained. For the 82 stars identified as variable, 
we give all necessary information needed to plan further phase dependent 
observations.
\end{abstract}

\section{Introduction}\label{introduction}
About 10\,\% of upper main sequence stars are characterised by remarkably 
rich line spectra. Compared to the solar case, overabundances are often 
inferred for some iron peak (e.g., Cr) and rare earth (e.g., Eu) 
elements, whereas some other chemical elements are underabundant (e.g., Sc). 
Some of these chemically peculiar (CP) stars also exhibit organised magnetic 
fields with a typical strength from a few hundreds up to few tens of thousands 
of Gauss. Chemically peculiar stars are usually subdivided into four groups: 
metallic line stars (CP1), magnetic Ap stars (CP2), HgMn stars (CP3), 
and He-weak stars (CP4) \citep{preston1974}. Chemical peculiarities are due 
to the diffusion of the chemical elements, resulting from the competition 
between radiative pressure and gravitational settling, which in magnetic 
stars is guided by the magnetic field, possibly in combination with the 
influence of a weak, magnetically-directed wind \citep[e.g.,][]{babel1992}.

Several CP stars also display photometric variability, which is either due 
to pulsation or caused by rotational modulation. Some CP1 stars, falling 
inside the instability strip, display for example $\delta$\,Sct pulsation 
\citep{kurtz1989}, while rapid oscillations characterise some CP2 (ro\,Ap) 
stars \citep{kurtz1982}. Except for CP1 stars, diffusion leads to the 
formation of surface spots which produce photometric variability with the 
same periodicity as the stellar rotation period. The light curves of those
stars can be fitted well by a sine wave and its first harmonic, thus allowing
recovery of the stellar rotation period with high precision.

Making use of photometric data obtained with the \stereo\ spacecraft, we
analysed the photometry of 1028 stars listed in the \citet{renson09}
catalogue of known or suspected CP stars. This work presents the results
obtained for the magnetic chemically peculiar stars (mCP stars: CP2 \& CP4).

NASA's twin \stereo\ spacecraft aim primarily to study the solar corona
in three dimensions and image Coronal Mass Ejections (CMEs) along the 
Sun-Earth line \citep{kaiser08} but its data can also be used to gather
high precision space photometry of background stars. As a matter of fact,
photometry extracted from data obtained by the Heliospheric Imager (HI) on 
board the \stereo-Ahead spacecraft \citep{eyles2009} has already been used to 
produce a catalogue of eclipsing binary stars (EBs) \citep{wraight2011}. The 
high photometric precision is ensured by the stability and accuracy of the 
calibration, as described by \citet{brown2009} and \citet{bewsher2010}.
In this work we use data obtained by both the Ahead \stereo\ spacecraft 
(\stereo/HI-1A) and the Behind spacecraft (\stereo/HI-1B) using a custom 
matched filter algorithm to extract the signals of variability. 

The \stereo/HI-1 data have a cadence of about forty minutes and cover in total 
four and half years, where each data-block, herein referred to as an epoch, 
from each spacecraft covers about twenty days with gaps of about one year. 
Given the fact that most of the photometric variability of mCP stars has a 
periodicity between several hours and a few days, \stereo\ photometry is 
perfectly suitable for the detection and measurement of this periodic 
signal.
\section{Observations and analysis}\label{obs}
\subsection{Characteristics of \stereo\ data}\label{data}
\begin{figure}
\begin{center}
\includegraphics[width=85mm,clip]{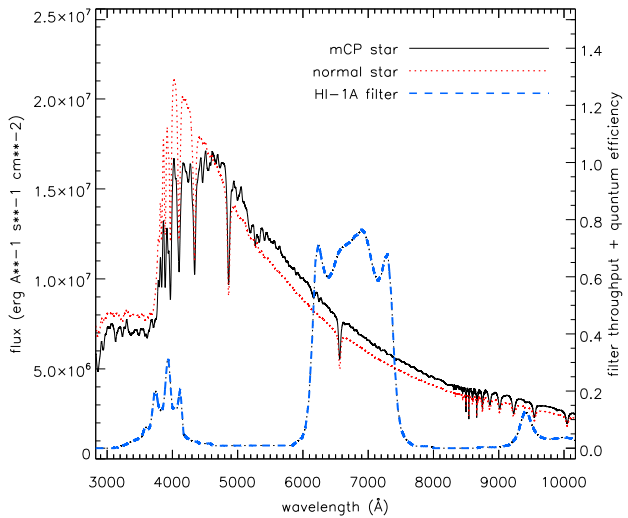}
\caption{Synthetic spectral energy distributions calculated with \Teff=8000\,K 
and \logg=4.0, and assuming classical mCP chemical peculiarities and a magnetic 
field of 10\,kG (full black line) and normal element abundances (dotted 
red line). The blue dashed line shows the HI-1A filter throughput, convolved 
with the CCD quantum efficiency, for which the scale is given by the y-axis 
on the right hand side of the plot.} 
\label{fig:filter} 
\end{center} 
\end{figure}

The \stereo/HI-1 images have a field of view of 20 degrees by 20 degrees. 
The images, collected with a 2k\,x\,2k pixel CCD, are binned on-board 2x2, 
returning to Earth images of 1k\,x\,1k with a pixel scale of 70\,\asec\ per 
pixel. All images use a single filter, with a spectral response mostly between 
6300\,\AA\ and 7300\,\AA, although the bandpass does allow some light through 
around 4000\,\AA\ and 9500\,\AA\ \citep{bewsher2010}. Figure~\ref{fig:filter} 
shows the HI-1A filter throughput, convolved with the CCD quantum efficiency, 
(dashed blue line) in comparison with synthetic spectral energy distributions 
(SED), calculated with \llm\ \citep{llm}, of two hypothetical stars of the 
same effective temperature (\Teff=8000\,K) and surface gravity (\logg=4.0); 
one SED was calculated assuming classical mCP chemical peculiarities and a 
magnetic field of 10\,kG (full black line), while the second SED was 
calculated assuming solar abundances and no magnetic field (dotted 
red line). It is important to notice that the amplitude of the variability, 
in the visible spectral range, of classical mCP stars decreases with 
increasing wavelength \citep{mikulaek2007}. For this reason, the window in 
the filter throughput around 4000\,\AA\ allows the \stereo\ photometry to be 
rather sensitive to the photometric variability of mCP stars.

As the primary purpose of the \stereo\ images is to observe coronal mass 
ejections (CMEs) along the Sun-Earth line, their field of view is centred 
14 degrees away from the Sun's limb. Over the course of an orbit, almost 
900,000 stars of 12th magnitude and brighter are imaged within 10 degrees 
of the ecliptic plane. The data very close to the brightest parts of the 
solar corona in the field of view can be affected by solar activity and 
this region was excluded from the analysis. 

\begin{figure}
\begin{center}
\includegraphics[width=85mm,clip]{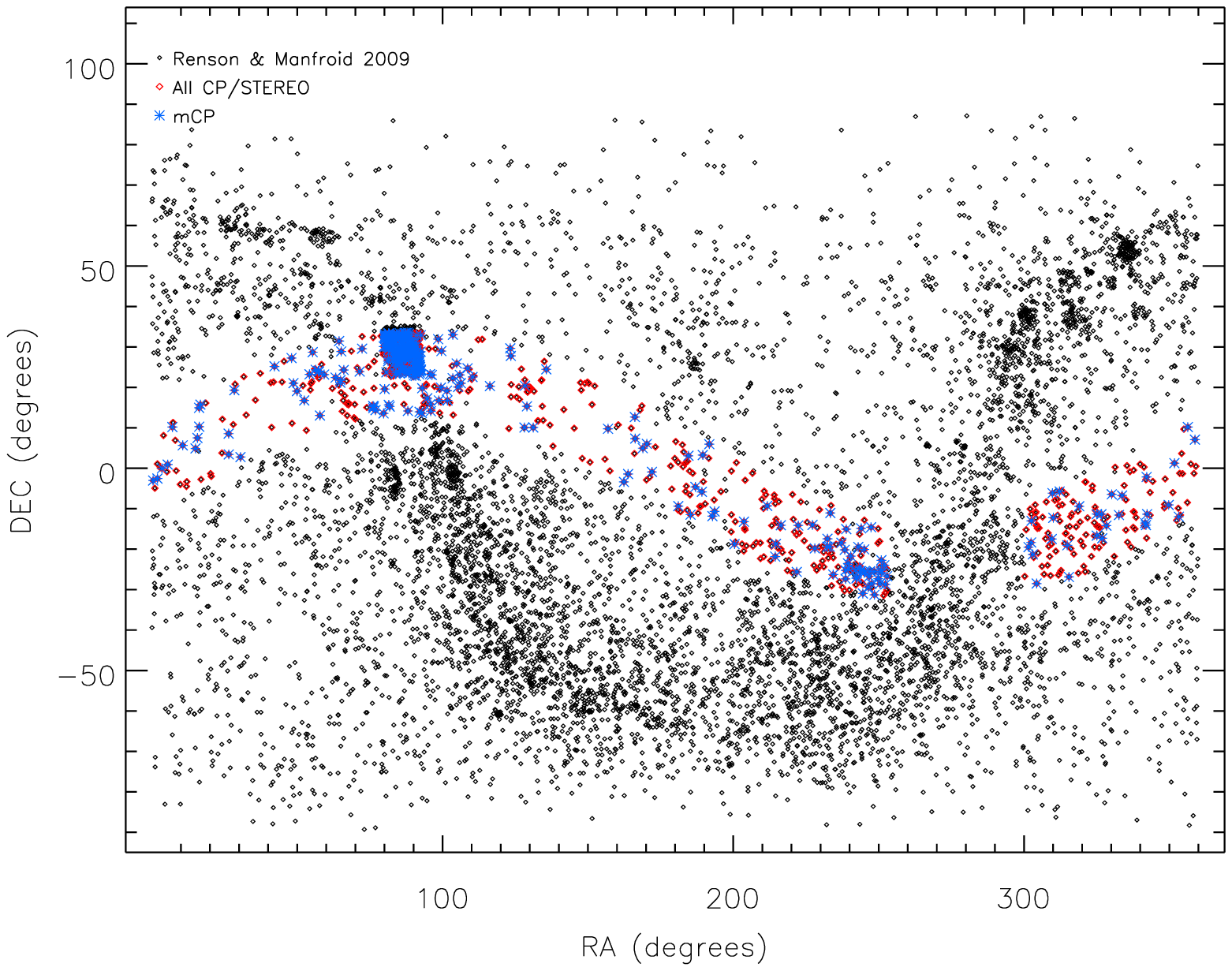}
\caption{Black points show the position in the sky of all stars listed in 
the \citet{renson09} catalogue. The red circles show the position of all
1028 CP stars present in the \stereo\ images. The blue asterisks show the 
position in the sky of the 337 mCP stars presented in this work. The clump 
of stars at RA$\sim$90\,deg corresponds to the CP stars identified by 
\citet{kharadze90}. Right ascension and declination are given in degrees.} 
\label{fig:skydistribution} 
\end{center} 
\end{figure}

Figure~\ref{fig:skydistribution} displays the position in the sky of all stars
listed in the \citet{renson09} catalogue, superimposed with the position of all
1028 analysed CP stars (red circles) and of the 337 mCP stars presented in 
this work (blue asterisks). Figure~\ref{fig:skydistribution} shows clearly 
that we discarded from the analysis all CP stars close to the densely 
populated Galactic center. We applied this cut-off because of the strong 
blending in the \stereo\ images caused by the large pixel scale.

\begin{figure}
\resizebox{4cm}{!}{\includegraphics[width=4cm,clip]{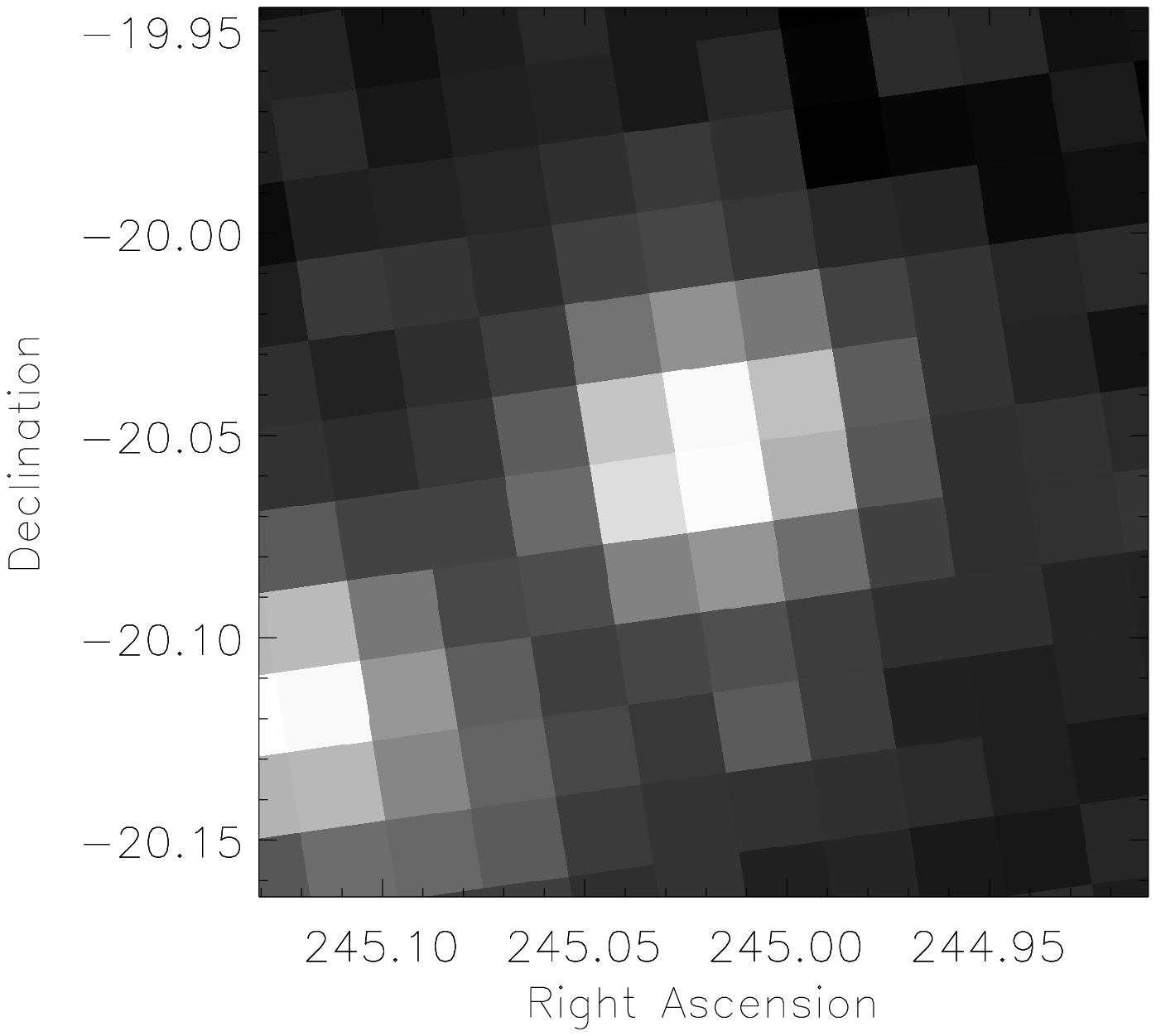}}
\hfill
\resizebox{4cm}{!}{\includegraphics[width=4cm,clip]{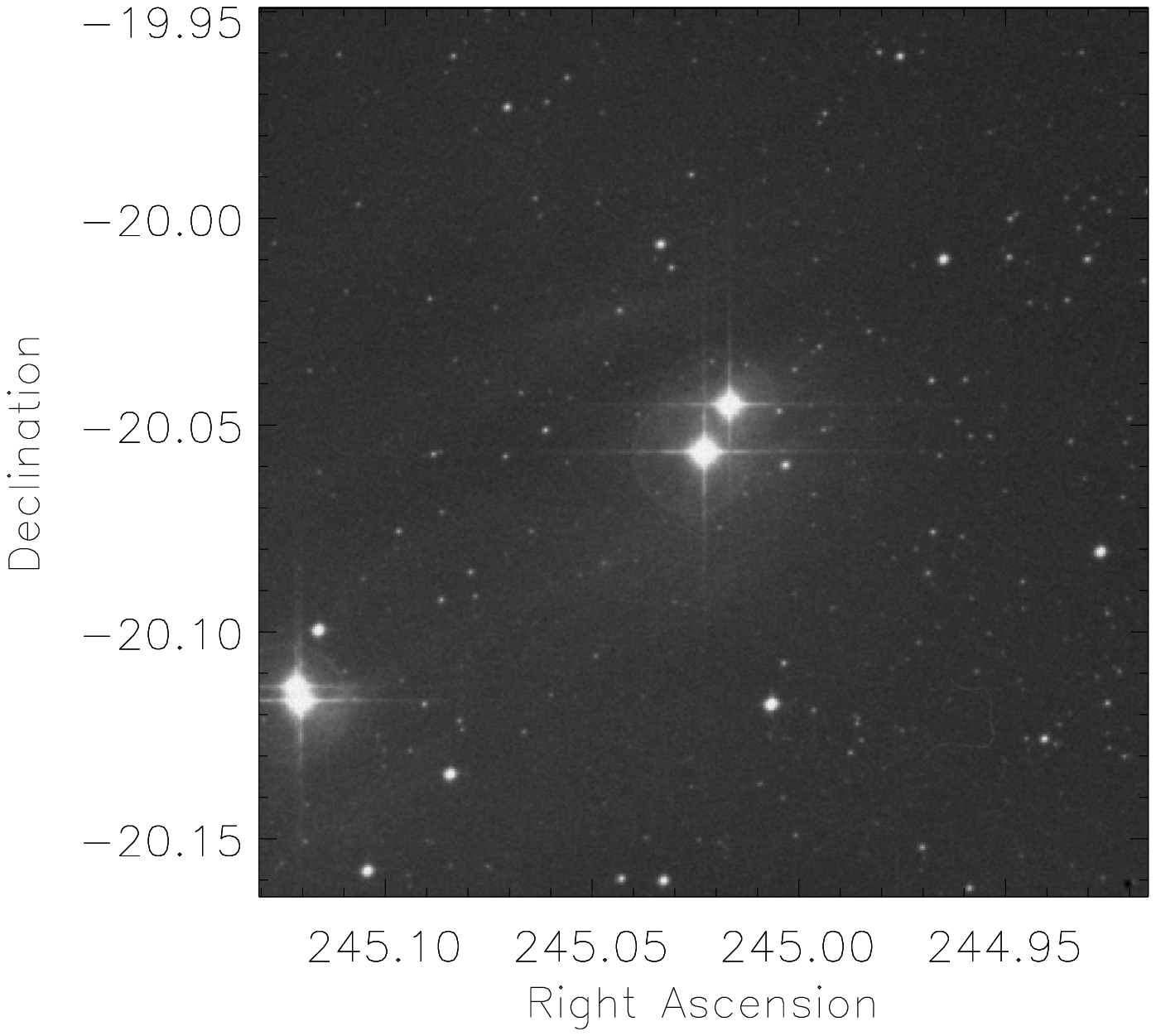}}
\caption{\stereo/HI-1A (left) and \textsc{AAO} \textit{R} band (right) views 
of a visual binary consisting of the CP2 star HD~147010 and its slightly 
fainter neighbour HD~147009, illustrating the severity of blending with 
the 70\,\asec\ pixel scale of the \stereo/HI-1 images.}
\label{blending}
\end{figure}

Figure~\ref{blending} shows the CP2 star HD~147010 as seen by \stereo/HI-1A 
in comparison with the image in the $R$ band of the same region of the sky 
taken from the Anglo-Australian Observatory (AAO). It is not possible 
from the \stereo\ photometry alone to determine which of the two central stars 
is the source of the observed variability. This same problem is also present 
for other variables, for which a confirmation of the variability of a certain 
star requires additional data. In the particular case shown in 
Fig.~\ref{blending}, the nearby star is HD~147009, a bright ($V$=8.06\,mag) 
star with neither chemical peculiarities nor any known variability, thus 
the likely source of the observed signal is HD~147010.

\begin{figure}
\begin{center}
\includegraphics[width=85mm,clip]{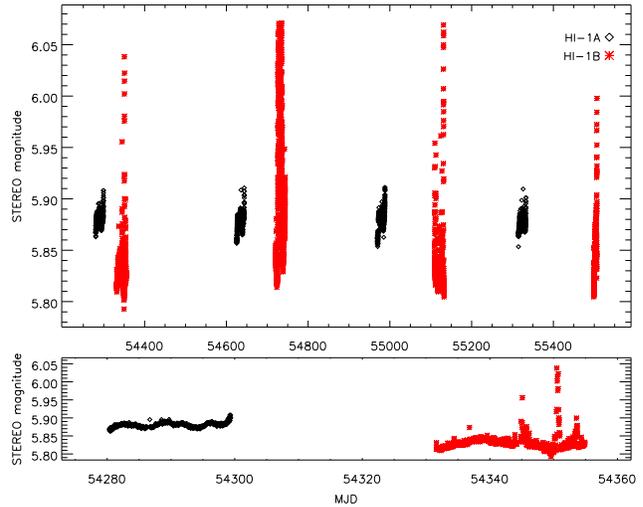}
\caption{Undetrended light curve of the CP2 star HD~74521. The black diamonds
indicate the \stereo/HI-1A data-set and red asterisks indicate the 
\stereo/HI-1B data-set. As the two satellites progress in their orbits, 
the interval between \stereo/HI-1A and \stereo/HI-1B observations increases. 
Systematic effects from the \stereo/HI-1B are evident. The first block of 
\stereo/HI-1A and \stereo/HI-1B data is enlarged in the bottom plot.}
\label{fig:fulllength} 
\end{center} 
\end{figure}

The \stereo-Ahead spacecraft is in an Earth-leading orbit with a semi-major 
axis of about 0.95\,AU, while the \stereo-Behind spacecraft is in an Earth 
trailing orbit with a semi-major axis of about 1.05\,AU. This results in stars 
remaining in the field of view of the \stereo/HI-1A imager for just over 
19\,days and in the field of view of the \stereo/HI-1B imager for just over 
22\,days. The images have a cadence of 40 minutes. Figure~\ref{fig:fulllength}
shows the light curve of the CP2 star HD~74521, illustrating the 
time separation between the different epochs of data, as well as some 
systematic effects which characterise a significant fraction of the 
\stereo/HI-1B data.

The \stereo/HI-1B data often exhibits sudden decreases in amplitude 
of all signals at a given time. These are attributed to errors in the 
background subtraction due to pointing changes resulting from micrometeorite 
impacts, as the \stereo/HI-1B imager is facing the direction of 
travel along the spacecraft's orbit. The severity of these events can range 
from a single observation point to days of erratic behaviour.  
Often we registered a marked decrease for some hours and then a return 
to the normality. Those systematics in the \stereo/HI-1B data can be easily
recognized as such by visual inspection as these artifacts are erratically 
shaped and, importantly, aperiodic, whereas the classic variability of mCP 
stars is highly periodic. 

The fact that the \stereo/HI-1A and \stereo/HI-1B satellites use slightly
different photometric apertures and that the background subtraction does not
take into account the number of stars in each field of view leads to a small
difference between the magnitude of the same star obtained with the two
satellites \citep{bewsher2010}. For this reason we normalised the 
\stereo/HI-1B data to the same level as the steadier \stereo/HI-1A data with 
a detrending algorithm, described in Sect.~\ref{analysis}.
\subsection{Data analysis}\label{analysis}
Before being passed to a custom matched filter algorithm for analysis, 
all light curves are subjected to a culling routine to remove some of the 
artifacts caused by Mercury and Venus passing through the field of view, as 
well as the more extreme pointing-related systematics present in the 
\stereo/HI-1B data, described in Sect.~\ref{data}. This routine removes 
all data points more than 4 standard deviations away from the weighted mean 
magnitude. Polynomial detrending is then carried out using a 4th order 
polynomial in order to remove residual trends that may remain from the 
flat fielding or other artifacts. This removes some variability that is 
long compared to the length of an epoch, thus periods longer than about half 
the length of an epoch are not expected to be highly reliable. With the
\stereo\ data, it is not possible to use differential photometry to avoid the 
detrending procedure, because trends are largely dependent upon the stars' 
location on the CCD, preventing differential photometry from being useful 
on a large scale. The \stereo/HI-1B data and also the succeeding epochs 
of \stereo/HI-1A data are also normalised to the same weighted mean magnitude 
as the first epoch of \stereo/HI-1A data.

The matched filter algorithm analyses a light curve in several stages, 
building model light curves and measuring the least-squared error of the model 
compared to the actual light curve. The model light curves consist 
of data points with the same time and errors as the real light curve. The 
matched filter algorithm works as follows:
\begin{itemize}
\item determine a best-fitting period. At this stage we adopted a sinusoidal 
shape, with an amplitude equal to three times the standard deviation, and 
produced the relevant periodogram.
\item Fine-tune the period. A precise period is the most important step in 
the process and without this the other characteristics will not be reliably 
determined.
\item Determine the amplitude of variability. This process is done before 
and after determining the best-fitting shape, with an amplitude resolution of 
5\,mmag.
\item Determine the shape of the variability. A selection of shapes 
consisting of sinusoidal variability with different harmonic signals 
overlaid are used in addition to shapes based upon box-like total eclipses, 
v-shaped eclipses and a composite of total eclipses with wide ingress and 
egress phases. The amplitude is recalculated after this stage and, if an 
eclipsing model was best-fitting, the algorithm also recalculates the duration 
and depth of the eclipses.
\item If the best-fitting amplitude is zero this means that the model 
finds the star to be constant. The period obtained at the very first step of the
algorithm is still returned.
\item If an eclipsing model is best-fitting and the amplitude non-zero, 
eccentricity and amplitude of secondary eclipses are checked for. In this
process, higher harmonics of the period are also checked.
\end{itemize}

This matched filter algorithm is processor-intensive and, depending on 
the number of data points (up to about 6000, currently), it may take from 5 to 
10\,minutes per star to process for a period range from 0.1 to 3.5\,days 
with a resolution of 0.005\,days ($\sim$7\,minutes) initially, then fine-tuned 
to 0.00005\,days ($\sim$4\,seconds) in a narrow search around the best-fitting 
period. Higher frequencies were excluded to avoid the Nyqvist frequency of 
about 15.625\,days$^{-1}$ (0.064\,days). This was the period range used in 
the initial search, however the program showed during testing a clear 
preference for finding the half-period harmonic of EBs, to the extent that 
it was specifically programmed to check at double and triple the periods 
initially found to see if these produced a better fit. A second search was 
then done with periods up to 10\,days using \stereo/HI-1A data only, as in 
some cases the \stereo/HI-1B data was of too poor quality to allow even strong 
variability to be detected. This particular upper limit of 10\,days was 
chosen to avoid running into systematic errors from the polynomial detrending 
at about one half of the length of an epoch.

The periodograms and light curves (phase-folded on the best fitting period) 
produced by the matched filter algorithm were visually inspected, primarily 
to extract from the sample the objects which appeared clearly constant.
Additionally, we classified as constant the stars which were so faint that 
any signal would be likely due to noise or if systematic effects were so 
extreme that the data were unusable, mostly stars at the very edge of the CCDs.
The same classification was given to the stars for which the lack of data
prevented the reliable detection of any variability. The list of those 133 
stars is given in Table~\ref{tab:junk}. 
The remaining 205 mCP stars that we did not consider definitely constant, or 
which merited further investigation on the basis of their features in the 
periodogram, were all individually examined in detail.

The first stage in this detailed analysis was to check the surrounding stars, 
within 8 pixels, as seen with the \stereo/HI-1A images, looking for signs of 
variability so that the risk of blending for each star could be understood.  
These stars have the potential for their signal and/or 
period to be affected by blending. The effects of blending can be broadly 
assigned to three different categories. {\it i}) Extreme cases where the 
target might not be the source of the variability and there is an equal 
chance the source may instead be a nearby star. {\it ii}) Severe cases where 
the target may be variable but there may be one or more other variables 
nearby and the true period has been distorted by the interference. 
{\it iii}) Minor cases where the target is clearly variable and no 
variability, or only vague variability, is seen in nearby stars and the 
only possible effect is a slight distortion of the period. Due to the 
\stereo/HI-1 pixel scale, most stars come into the latter category to some 
extent but the accuracy of the photometry precludes adjacent stars more 
than four magnitudes fainter from having a measurable influence. Thanks also
to the different photometric apertures between the two satellites, the 
presence of a difference between the variability extracted from the 
\stereo/HI-1A and the \stereo/HI-1B data also indicates the presence of 
blending. This method proved to be particularly helpful in assigning the 
source of variability for some blended stars. The 8 pixel range was chosen 
partly to be conservative, as blending effects have not been observed from 
such a distance, and partly to help trace whether a star closer than this to 
the target is genuinely variable or itself blended with a more distant 
neighbour. 

The second stage in the detailed examination was an inspection of the 
detrended light curve of the target, using both \stereo/HI-1A and \stereo/HI-1B 
data, where available. In particular, we examined the severity of artifacts 
present in the \stereo/HI-1B data, deciding carefully where to ``clean" the
light curves to avoid obscuring or distorting a potential signal. For 
systematic effects, most commonly artifacts of the polynomial detrending, 
the severity can range from introducing a completely false signal to a 
small risk of distorting the period of an existing signal. 

To attempt to remove remaining systematics, a 7th order polynomial was applied 
after the initial culling of outliers. This was sufficient to remove most of 
the effects of planetary incursions by Mercury and Venus. In extreme 
cases, it was necessary to remove entire epochs or even completely exclude 
\stereo/HI-1B data from the analysis, however some stars only needed a few 
clear outlying points removing and most only required minor cleaning, 
typically for small numbers of events due to pointing errors as described 
in Sect.~\ref{data}.

The final stages of the detailed analysis were undertaken with
Peranso\footnote{\tt http://www.peranso.com}. We undertook the period search 
with two algorithms, to cross-check each other and avoid duplicating 
weaknesses. In each case, we searched periods between 0.1 and 10\,days, 
although in a few individual cases a search was made outside this range. The 
Lomb-Scargle method \citep{scargle82} was employed in the period domain and 
the phase dispersion minimisation (PDM) method \citep{stellingwerf78} was 
employed in the frequency domain. We then examined the significant features 
in the periodogram produced with each method to extract the most likely 
period, its uncertainty and the epoch of the first maximum in the \stereo\ 
light curve.
\section{Results}\label{results}
Following the procedure described in Sect.~\ref{analysis}, we compiled 
Table~\ref{tab:const}, listing the 122 stars we classified as constant after 
the detailed individual analysis, and Table~\ref{tab:variable}, listing the 82 
stars we classified as variable. Those tables, together with 
Table~\ref{tab:junk}, present the main results of this work on the mCP stars. 
Each table lists the star name and the \citet{renson09} identification number 
in the first two columns. The third and fourth columns list the coordinates 
of each star (in deg), while the average magnitude in the $V$ band is given 
in the fifth column. Column number 6 lists the spectral classification and 
chemical peculiarity given by \citet{renson09}, while the seventh column 
lists the CP classification. The starting point for the critical assessment 
of the literature about the CP classification of our programme stars was the 
Catalogue of Ap, HgMn and Am stars by \citet{renson09}. They collected all 
objects which were reported as ``peculiar" in the literature and list an 
approximate spectral type. Because at least classification resolution 
spectroscopy is necessary to establish the true nature of a peculiar object 
\citep[e.g.][]{paunzen2011}, we used the relevant extensive catalogue by 
\citet{skiff2010} to verify all spectral types. If contradicting 
classifications were found, a question mark was set. The CP classification 
given in these Tables does not take into account the findings of this paper 
and is based only on the information given in the literature. 

As mentioned in Sect.~\ref{analysis}, Table~\ref{tab:junk} lists all stars that
were immediately classified as constant, but it also includes stars for which 
the photometry was clearly affected by systematic effects and/or by blending, 
making the detection of any periodicity impossible. In particular, 
in Table~\ref{tab:junk}, for all stars brighter than $V\sim$5\,mag, the 
presence of systematic effects prevented us from detecting any periodic signal. 
Similarly, unless an exceptionally strong signal was present, the quality 
of the data did not allow us to conclude anything about stars fainter than 
$V\sim$10.5\,mag. As a consequence, all stars listed in Table~\ref{tab:junk}, 
with a magnitude between about 5 and 10.5 are either constant, or variable 
with a periodicity shorter/longer than 0.1/10\,days, or variable with an 
amplitude below our sensitivity.

As mentioned before, Table~\ref{tab:const} lists the stars that after
individual analysis we either classified as constant, or for which
systematics or blending prevented us from detecting reliable signals. 
Column 10 gives the particular reason for a certain star to be listed in this
table: C when we concluded the star was constant\footnote{This means that the
star is either constant, or variable with a periodicity shorter/longer than 
0.1/10\,days, or variable with an amplitude below our sensitivity, or
pole-on.}, B and S when respectively blending and systematics were too 
severe to allow the detection of a reliable signal. In a few cases we detected 
rather weak signals, with a small significance (see below). For those stars 
we added a W in column 10 and the period (in days) with its uncertainty (in 
days) respectively in columns 8 and 9. Those periods have to be taken with 
caution. In column 11 we finally listed the period found in the literature.

In Table~\ref{tab:variable} we list the objects for which we found and 
measured a genuine period (in days), listed in column 8, together with the 
period uncertainty (in days) in column 9 and the MJD of the epoch of the first
recorded maximum in column 10. As shown in columns 8 and 9, in some cases we
detected more than one significant period. As for Table~\ref{tab:const}, in 
the 11th column we included a remark indicating the possible presence of 
blending (B) or systematic (S) effects which could have slightly distorted 
the given value of the period. For those stars, marked with B and/or S, the 
effects of blending and systematics were not large enough to doubt the 
source of the variability. In column 11, we put an additional asterisk ``*" to 
highlight the stars for which we obtained an exceptionally strong signal. 
Column 12 lists the period found in the literature.

For each star in Table~\ref{tab:variable} we produced a periodogram
(Fig.~\ref{fig:allperiodograms}) and a phase-folded light curve
(Fig.~\ref{fig:alllightcurves}). The periodograms were produced using one of 
the outputs of the PDM method, called $\Theta$, defined in 
\citet{stellingwerf78}. Given the large number of photometric points, 
$\Theta$ gives a direct indication of the significance of a certain period 
\citep[see Eq.~13 in][]{stellingwerf78}, which we then defined as the 
deviation from the median value of $\Theta$ in units of standard deviations. 
All periods obtained for the stars that were genuinely 
variable have a significance larger than 7, where the maximum value is 
reached for HD\,19832 (SX\,Ari) with a significance of about 32. In 
Table~\ref{tab:variable} we decided not to include the significance of the
period because it is not the only factor which played a role in deciding 
whether a certain period was real or not; external factors, blending in 
particular, had to be taken into account, making the statistical 
significance misleading.

\begin{figure}
\begin{center}
\includegraphics[width=85mm,clip]{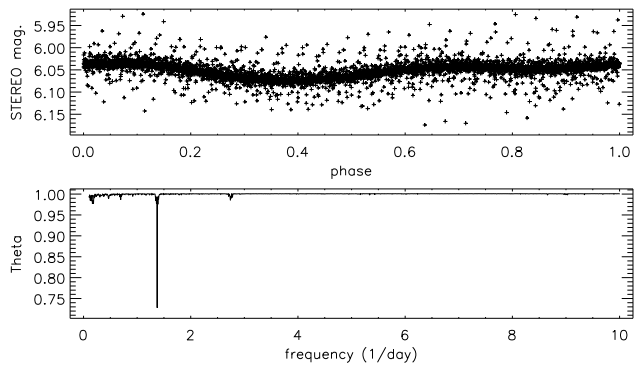}
\caption{Top: light curve obtained for the CP2 star HD\,19832 phase-folded 
with the most significant period of 0.7279\,days. The periodogram is shown in
the bottom panel. This mCP star is clearly variable and is listed in
Table~\ref{tab:variable}.}
\label{fig:variable} 
\end{center} 
\end{figure}

\begin{figure}
\begin{center}
\includegraphics[width=85mm,clip]{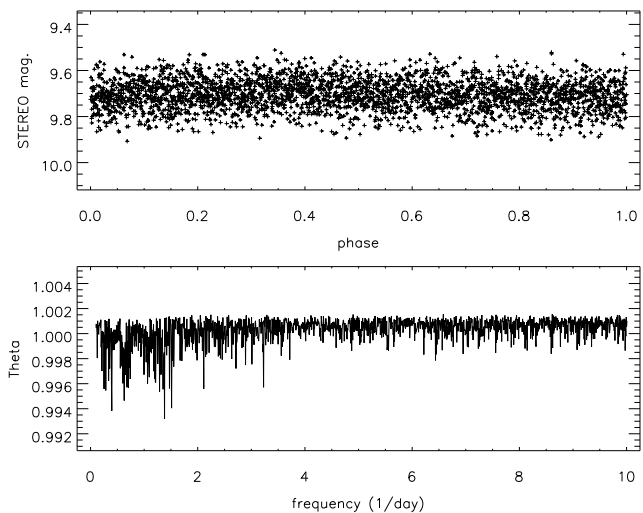}
\caption{Same as in Fig.~\ref{fig:variable}, but for HD\,149228. For this mCP
star, listed in Table~\ref{tab:const}, we did not obtain any significant 
period, meaning that the star is either constant or has a rotation period 
longer than 10\,days. In this case the light curve is phase folded on the 
period given by the most significant peak in the periodogram (1.5066\,days).}
\label{fig:constant} 
\end{center} 
\end{figure}

\begin{figure}
\begin{center}
\includegraphics[width=85mm,clip]{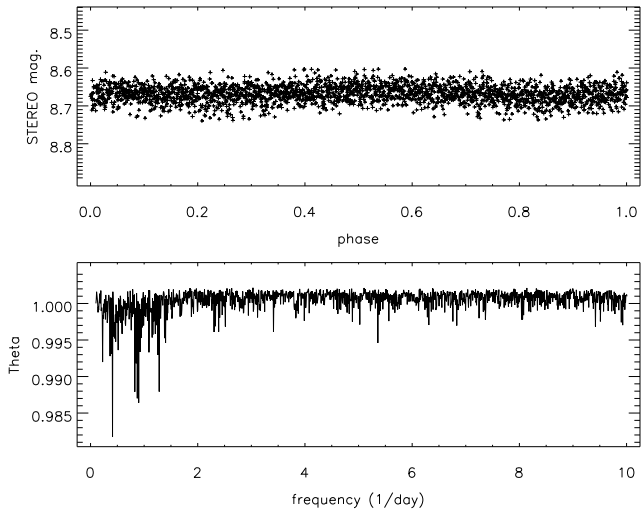}
\caption{Same as in Fig.~\ref{fig:variable}, but for HD\,150035 and 
phase-folded on a period of 2.3389\,days. This star is a border-line case 
(see text for more details).}
\label{fig:middle} 
\end{center} 
\end{figure}

Figure~\ref{fig:variable} shows the phase-folded light curve (top) and the
periodogram (bottom) for HD\,19832 (SX\,Ari), for which the derived period 
presents the strongest signal in our sample. In Fig.~\ref{fig:constant} and 
in Fig.~\ref{fig:middle} we show the periodogram and the phase-folded light 
curve for HD\,149228 and HD\,150035, respectively a star we classified as 
constant and a star for which we found a weak signal (see 
Table~\ref{tab:const}). HD\,150035 is probably the most border-line case in 
our sample. The period stated in Table~\ref{tab:const} has a significance 
below 7 and it does not match the previously known period given by 
\citet{catalano1998}, although this is a rather bright star and neither 
blending nor systematics affect the photometry, as shown by the neatness of the
phase-folded light curve. It is also possible that the amplitude of the
variability is below the sensitivity of the instrument.

A handful of stars listed in Table~\ref{tab:const} and \ref{tab:variable}, 
for which we measured a period, are the primary component of a spectroscopic 
binary system, e.g. HD\,68351. For each of them, we did not find any relation 
between the orbital period of the binary system, given by \citet{sb9}, and 
the period obtained from the \stereo\ photometry.

\begin{figure}
\begin{center}
\includegraphics[width=85mm,clip]{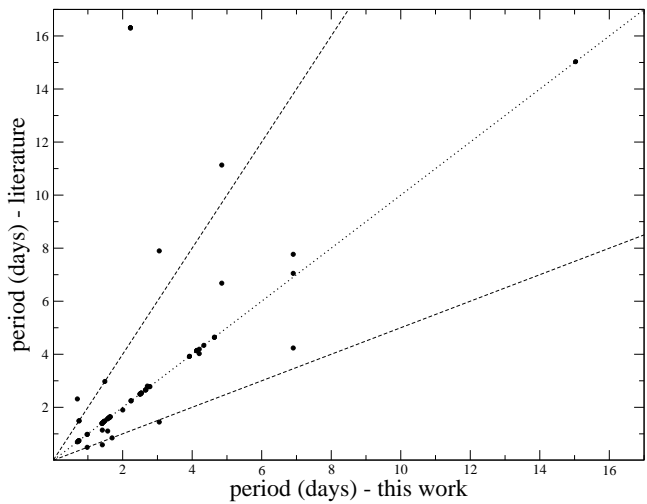}
\caption{Comparison between the periods of mCP stars derived from the \stereo\ 
data and present in the literature. The dotted line shows the one-to-one
relationship, while the dashed lines indicate the relationships relative to 
twice and half of the literature period.}
\label{fig:periodlit} 
\end{center} 
\end{figure}

Figure~\ref{fig:periodlit} shows a comparison between the periods derived from
the \stereo\ data and those found in the literature. We register a general good
agreement, except for the cases when the \stereo\ period is a harmonic of the
one given in the literature, as for HD\,12447\footnote{This star is also a
blend with the possible CP1 HD\,12446.}. Table~\ref{tab:variable} shows
that this also happens quite often among the various periods present in the
literature. The largest discrepancy is obtained for HD\,111133 where the 
literature period is about 7.3 times larger than the one we obtained. In 
this case the published $\sim$16\,days period is hardly recognisable in our 
data; a weak peak in the periodogram at a period just below 16\,days, was 
found as in this case we extended our period search to more than 10\,days to 
look for the period given in the literature. In our periodogram, the 
$\sim$2.2\,days period gave the strongest signal, with a significance of 
about 8.
%
\subsection{Notes on individual stars}\label{notes}
The stars HD\,965, HD\,142250, HD\,209051, and HD\,216018, confirmed as 
magnetic by previous spectropolarimetric measurements 
\citep[][, respectively]{bychkov2009,kudryavtsev08,romanyuk08}, are listed 
among the constant stars in Table~\ref{tab:junk}. For those objects our 
periodogram did not show any significant peak in the 0.1--10\,days period 
range, meaning that their rotation period is likely to be longer than 
10\,days. For these four stars the rotation period is not known.

For the stars HD\,26571 and HD\,134214, listed in Table~\ref{tab:junk},
\citet{catalano1998} and \citet{renson01} gave very different periods. For 
HD\,26571, \citet{catalano1998} gave a period of 1.06\,days, while
\citet{renson01} gave a period of 15.749\,days. Our periodogram did not show
any significant peak around the 1\,day period, therefore we presume 
that the 15.749\,days period, given by \citet{renson01} is the correct one.
\citet{adelman2008} confirmed the longer period and refined it to 
15.7505\,days. For HD\,134214, \citet{catalano1998} gave a period of 
248\,days, while \citet{renson01} gave a period of 4.15\,days. Our 
periodogram did not present any significant peak around 4\,days, meaning 
that the longer period is more likely to be the correct one.

For the star HD\,27295, listed in Table~\ref{tab:junk}, \citet{catalano1998}
reported a rotation period of 4.42\,days, but our periodogram did not 
present any statistically significant period and also the detrended light 
curve did not show any sign of variability. Since the star has been confirmed
to be magnetic, it is possible that the rotation period is longer than
10\,days. \citet{samus09} listed the star HD\,17471 (see Table~\ref{tab:junk}) 
as a variable of the $\alpha$2\,CVn type, without giving any period. Our light 
curve for this star is distorted by systematic effects, therefore we cannot
provide a measurement of the rotation period. 

Except for the very bright and very faint stars, for which systematic 
effects prevented us from detecting the presence of genuine variability, for
the remaining objects listed in Table~\ref{tab:junk} we did not detect any
significant period. For those stars there is no spectroscopic and/or 
polarimetric confirmation of their magnetic nature. We believe that most of
those objects are either CP1 or chemically normal stars, while it is possible
for some of them to be mCP stars with a rotation period longer than
10\,days, or variable with an amplitude below our sensitivity.

For the mCP stars HD\,47103 and HD\,148321 \citep[see][]{bychkov2009}, listed 
in Table~\ref{tab:const}, we did not find any significant variability, 
although the data did not present any systematic effect and the photometry 
was not affected by blending. Most likely the rotation period of these 
two stars is longer than 10\,days. Blending prevented us from detecting any 
reliable and significant variability for the mCP stars 
\citep[see][]{bychkov2009} HD\,22374, HD\,23387, and HD\,51688, also listed in
Table~\ref{tab:const}.

For the magnetic stars HD\,224926, HD\,125248, and HD\,150035 
\citep[see][]{bychkov2009}, listed in Table~\ref{tab:const}, we identified a
periodicity with a significance between 5 and 7, where for the first and second
stars respectively, blending and systematic effects prevented us from 
obtaining a period convincing enough to include them among the certainly 
variable mCP stars. The period we measured for HD\,125248 is in agreement 
with the one previously reported by \citet{renson01}.

\citet{renson01} listed a rotation period of 1.563\,days for 
GSC\,00742-02169, while we obtained a significant period at 1.4828\,days, 
which might have been distorted by blending. In Table~\ref{tab:const} we
listed the stars HD\,10809, HD\,250027, and HD\,146998 as constant, although a
previous period was reported by \citet{kraus}, \citet{samus09}, and
\citet{catalano1998}, respectively. For both HD\,10809 and HD\,146998 we have 
not found any significant peak in the periodogram around the previously 
reported period values. For HD\,250027 the period given by \citet{samus09} of
about $\sim$20\,days is out of our detectability window.

For the stars HD\,1758, GSC\,02390-00208, and AAO+27\,25, listed in
Table~\ref{tab:const}, we identified a peak in the periodogram which is
considered not significant enough to classify these objects as certainly
variable. We came to the same conclusion for the stars HD\,23850, HD\,23964, 
HD\,242692, GSC\,02403-00597, AAO+27\,185, HD\,39865, AAO+30\,338, HD\,48953,
BD+23\,1580, GSC\,01398-00532, HD\,118054, HD\,138426, HD\,139160, HD\,144748,
HD\,151941, HD\,215766, and HD\,215913, but for those objects blending and/or 
systematic effects might have distorted the measured period. For these stars,
there is no spectroscopic and/or polarimetric confirmation of their magnetic
nature. HD\,196470 is a magnetic ro\,Ap star for which blending prevented
detection of any significant peak in the periodogram. Similarly, HD\,206088 
is a mCP star for which systematic effects did not allow us to measure any 
reliable period. For both HD\,196470 and HD\,206088 it is possible that their 
rotation period is longer than 10\,days, or variable with an amplitude below 
our sensitivity.

For the remaining stars listed in Table~\ref{tab:const} there is no 
spectroscopic and/or polarimetric confirmation of their magnetic nature and 
most of those objects are likely to be either CP1 or chemically normal stars, 
with a minority of mCP stars with a period longer than 10\,days, or variable 
with an amplitude below our sensitivity.

About half of the stars we classified as variable and listed in
Table~\ref{tab:variable} are known mCP stars and for some of them the 
rotation period has been previously measured. For those stars the results 
provide either a confirmation or a refinement of the previously known period, 
with the addition of a period uncertainty and of an epoch of maximum 
brightness, which is crucial information to plan spectroscopic observations 
aiming to perform, e.g., Doppler imaging. It is important to notice that 
blending and/or systematic effects, reported in column eleven of 
Table~\ref{tab:variable}, could have slightly distorted the given period. 
For a few stars we registered also the presence of more than one significant 
period in the periodogram, which we reported in Table~\ref{tab:variable}.

For the remaining 48 stars, listed in Table~\ref{tab:variable}, the results
represent the first measurement of their rotation period. In particular, very 
little is known for most of them and the clear detection of rotational
variability is strongly suggestive of the presence of a structured magnetic 
field, chemical peculiarities and surface spots, although some of them might 
be CP3 stars, and therefore not magnetic. It would be extremely valuable
to obtain spectropolarimetric observations for those stars, which could be
quickly performed with instruments such as \espa\ at the Canada-France-Hawaii
Telescope (CFHT). In particular, given the average magnitude of those stars 
and because their magnetic field (if present) should be of the order of a few
hundreds of Gauss, their detection will not require spectra with a large
signal-to-noise ratio and therefore, such observations could be performed 
with short exposure times.

For the stars HD\,43819 and HD\,130559 we reported values of the rotation 
period longer than 10\,days, adopted as maximum for the matched-filter 
algorithm. For HD\,43819 we looked intentionally for a period 
around 15\,days, because \citet{renson01} reported a period of 15.03\,days. On
the other hand, for HD\,130559 the possible presence of a period
longer than 10\,days appeared from the undetrended light curve,
nevertheless, this value of the period has to be taken with caution.
\section{Discussion}
For each star, listed in Table~\ref{tab:variable}, we compiled Johnson $UBV$, 
Str\"omgren, and Geneva photometry from the General Catalogue of Photometric 
data \citep{mermilliod97} and the additional literature to derive the 
effective temperature on the basis of the calibrations given by 
\citet{netopil08}. These calibrations can be applied because they are 
specifically tuned for the different types of CP stars. The final \Teff, given 
in column 15 of Table~\ref{tab:variable}, is the average \Teff\ obtained 
calibrating the different colors, while the standard deviation and the number 
of averaged temperatures are given in parentheses. When available, we adopted 
the spectroscopic \Teff\ listed in \citet{netopil08}, and these 
cases are indicated with a ``99" instead of the number of averaged 
temperatures. To each star, with a determined \Teff, we associated a further 
fixed uncertainty of 500\,K for the CP2 stars and of 700\,K for the CP4 stars, 
as proposed by the reference above 
\citep[see also the discussion in][]{landstreet07}. For the following analyses 
we then adopted the largest uncertainty between the standard deviation and 
the fixed uncertainty.

For some stars later than spectral type A0 without Str\"omgren photometry, 
we were not able to determine individual reddening values based on the 
available photometry \citep[see the discussion in][]{netopil08}, hence the 
determination of \Teff\ would be rather erroneous. In order to also obtain 
temperatures for these (CP2) objects, we made use of the spectral energy 
distribution (SED) fitting tool by \citet{robitaille07}, which to some extent 
allows the extinction to be set as a free parameter. As input data we used the
available $UBV$ photometry or Geneva measurements transformed to $UBV$ using 
the calibration by \citet{harmanec01}, in combination with 2MASS data 
\citep{cutri03}. Although the $U$ magnitude is important for a proper fit 
of the energy distribution, the SED fitting was also applied to four objects 
without this filter information using $B$ and $V$ data taken from the 
ASCC-2.5\,V3 catalogue \citep{kharchenko01}. However, since for these stars a 
distance is available, a restriction of extinction was made using the three 
dimensional extinction maps by \citet{arenou92} in combination with the dust 
maps by \citet{schlegel98}. Since the anomalous colours of CP objects have to 
be taken into account, a correction of the SED results is probably necessary. 
Therefore, we also processed the CP2 ``standard" stars given in 
\citet{netopil08} in a similar way and determined the relation:
\begin{equation}
\label{qfit}
T_{\mathrm{eff}} = 1032(167)\,[K] + 0.873(16)\,[K]\,\,T_{SED},
\end{equation} 
with a correlation coefficient $R=0.990$ using in total 62 objects, valid 
for the temperature range 7500$-$15000\,K.

The obtained (SED) reddening was finally applied to the available photometry 
of our programme stars to obtain an additional second temperature estimate, 
which was averaged. These results are in good agreement with the 
known spectral types, and are indicated with ``50" in column 15 of 
Table~\ref{tab:variable}, instead of the number of averaged temperature 
determinations. We could not apply this method to stars for which 
neither a parallax nor a $U$ magnitude were available. These objects were 
detected by \citet{kharadze90}, the only available reference for most of them.

For the stars for which a parallax is present in the HIPPARCOS catalogue 
presented by \citet{leeuwen}, we determined the luminosity (\logl) on the 
basis of the Johnson $V$ magnitude, the bolometric correction given by 
\citet{netopil08}, and the interstellar reddening $E(B-V)$ determined via 
the different photometric systems \citep[see][for details]{netopil08} or 
via SED fitting, using a total-to-selective absorption ratio of $R=3.1$. 
Table~\ref{tab:variable} lists in column 16 the derived luminosities 
with the relative uncertainties given in parentheses.

For all stars for which we derived \Teff\ and luminosity, we determined 
stellar mass and fractional age ($\tau$ - fraction of main sequence 
lifetime completed) using the evolutionary tracks for solar metallicity 
given by \citet{schaerer}. Table~\ref{tab:variable} lists in columns 17
and 18 respectively the derived masses and fractional ages with the 
relative uncertainties in parentheses. In Table~\ref{tab:variable}, the 
stars for which the mass is marked with a ``Z" or ``T" lie below the zero 
age main sequence (ZAMS) or above the terminal age main sequence (TAMS). 
In these cases the mass was estimated as the star would lie directly on the 
ZAMS or TAMS and the fractional age has then been set to 0 (ZAMS) or 1 (TAMS), 
without uncertainty. Stars for which a temperature is available but no 
parallax, were placed in the middle of the main-sequence band to obtain at 
least a rough mass estimate; for these objects no luminosity and fractional 
age is given in Table~\ref{tab:variable}. 

\citet{bagnulo06} showed that for field stars important age uncertainties 
arise from the use of isochrones with an inaccurate metallicity, especially 
for stars in the first half of their main sequence lifespan. The introduction 
of the fixed uncertainties on \Teff\ of 500\,K for CP2 and 700\,K for CP4 
stars alleviates this problem. On the other hand, the use of tracks with an 
inaccurate metallicity has a much smaller impact on the derived masses.

The uncertainties given in Table~\ref{tab:variable} for the fractional ages 
are quite uniform in $\tau$, with an average of $\sim$0.2. \citet{landstreet07} 
compared the uncertainties they obtained for the fractional ages for a set 
of mCP stars in open clusters with the ones given by \citet{kb} in a set of 
field mCP stars. \citet{landstreet07} concluded that the uncertainties 
given by \citet{kb} were probably underestimated, due to the small 
uncertainties adopted for \Teff. In this work we adopt temperature 
uncertainties similar to those used by \citet{landstreet07}, making our 
uncertainties on the fractional ages more realistic compared to those of
\citet{kb}.

\begin{figure}
\begin{center}
\includegraphics[width=85mm,clip]{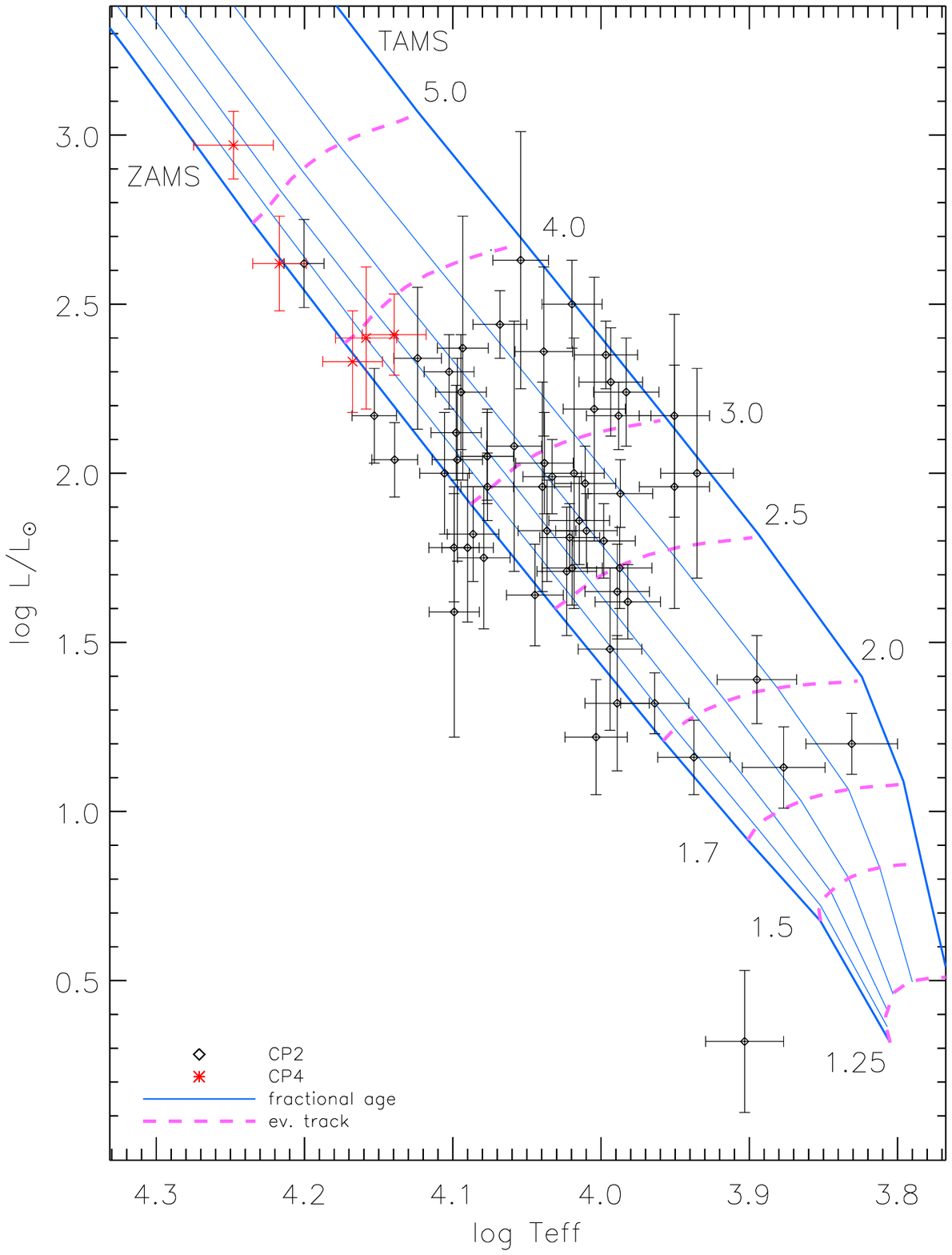}
\caption{Hertzsprung-Russell diagram for the CP2 (black diamonds) and CP4 (red
asterisks) stars listed in Table~\ref{tab:variable} and for which we
derived both \Teff\ and \logl. The blue lines represent the lines of equal
fractional age \citep{schaerer}. The ZAMS and TAMS are
indicated in the plot and highlighted with a thicker line. The dashed violet
lines represent the main sequence evolutionary tracks used to derive the 
stellar masses. The stellar mass, relative to each track, in units of solar
masses, is indicated in the plot. Both isochrones and evolutionary tracks are
for solar metallicity.}
\label{fig:hr} 
\end{center} 
\end{figure}

Figure~\ref{fig:hr} shows the position in the Hertzsprung-Russell (HR) 
diagram for all stars listed in Table~\ref{tab:variable} and for which we
derived both \Teff\ and \logl. Figure~\ref{fig:hr} shows also the evolutionary
tracks \citep{schaerer} adopted to derive the stellar masses
and fractional ages. As expected, the CP4 stars are on average hotter and more
luminous than the CP2 stars. Within the uncertainties, the position of the 
stars in the HR diagram is well inside the limits given by the ZAMS and TAMS, 
except for HD\,107452, which lies more than 2$\sigma$ below the ZAMS and is 
likely connected to a problem with the adopted distance. HD\,107452 is one of 
the closest objects of our sample, according to Hipparcos parallaxes. However, 
it is part of a close visual double star, probably influencing the parallax 
measurements. As a matter of fact, using the distance of 159\,pc given by 
\citet{gomez98}, which is more than twice of the Hipparcos one, the star 
would fall close to the ZAMS.

In Table~\ref{tab:variable} five stars (HD\,142301, HD\,142990, HD\,145501,
HD\,146001, and HD\,147010) belong to the Upper~Sco association and are 
present in the work by \citet{landstreet07}, who derived \Teff, \logl, 
\M, and $\tau$ for those stars. The agreement obtained for \Teff, \logl,
and \M\ is remarkably good, being almost always less than 1$\sigma$ away from
the previous value, while we notice relevant discrepancies for the fractional 
ages. We derived the fractional ages without considering the possible 
membership of open clusters or associations, while \citet{landstreet07} used 
this information to nail down the fractional age, sometimes to a 
few \%. This shows clearly the advantage of using open cluster stars, for 
which the age is precisely known, to study stellar evolution. However, we 
have to note that the individual Hipparcos distances for the stars in the 
Upper~Sco association are on average 10\% larger than the mean distance 
adopted by \citet{landstreet07}. This fact together with the adoption of a 
different temperature calibration (especially for CP4 objects) leads to a 
shift to larger fractional ages.

\citet{kb} compiled the values of the average quadratic longitudinal field,
rotational period, and evolutionary status for about 200 mCP stars. From the 
point of view of the evolution of stellar rotation, their results indicate
evidence for a increase of the stellar rotational period with increasing age, 
as a consequence of the conservation of angular momentum, which was previously
shown also by \citet{north98}. Unfortunately, our sample is too small and in 
particular biased towards shorter periods to be able to confirm or disprove 
\citet{kb}'s findings. Clearly, a thorough analysis of the evolution of the 
rotation period in mCP stars requires a large sample of open cluster stars.

In column 13 of Table~\ref{tab:variable} we added also the values of the 
average quadratic effective magnetic field, in Gauss. Those values have been 
taken from \citet{bychkov2009}, \citet{kudryavtsev08}, and \citet{romanyuk08}. 
When available, in column 8 of Table~\ref{tab:junk}, in column 12 of 
Table~\ref{tab:const}, and in column 14 of Table~\ref{tab:variable} we added 
the average projected rotational velocity (\vsini) values, put together on the 
basis of the compilation by \citet{glebocki05}. Newer observations were 
included from catalogues and papers found in the CDS/Simbad databases. 
Values of upper and lower limits as well as uncertain ones were discarded 
resulting in a few thousands of individual data points. The averages were 
calculated using the published errors as reciprocal weights. If no errors 
were listed, we set it to 15\%, which is a widely used value for such 
measurements.

\begin{figure}
\begin{center}
\includegraphics[width=85mm,clip]{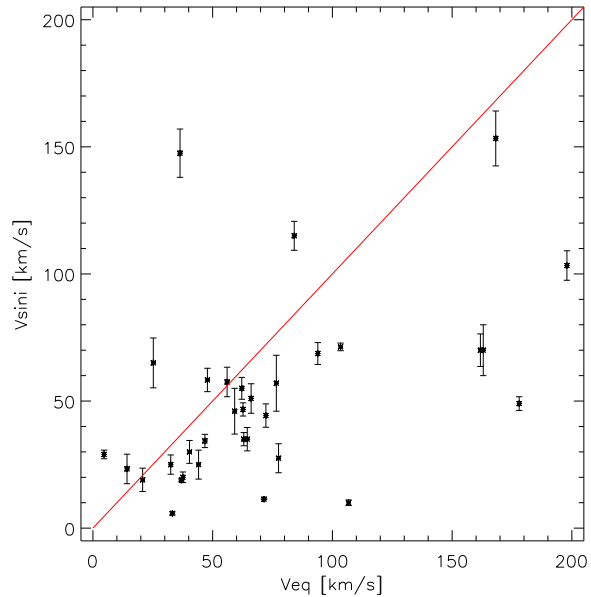}
\caption{Comparison between the observed \vsini\ and the computed \veq\ for 
the stars listed in Table~\ref{tab:variable}. The adopted \vsini\ uncertainty 
is also listed in Table~\ref{tab:variable}. The continuous line is the 
one-to-one relationship.}
\label{fig:vsini} 
\end{center} 
\end{figure}

As a test for the validity of the determined effective temperatures,
luminosities, and rotational periods we calculated the equatorial velocities
(\veq) from the formula of the oblique rotator model 
\citep[see e.g.][]{north98}:
\begin{equation}
\label{oblique model}
V_{eq} [km\,s^{-1}]=50.6\,\,R[R_{\odot}]\,\,/\,\,P[days]
\end{equation} 
where we calculated the stellar radii from \Teff\ and \logl.
Figure~\ref{fig:vsini} shows the comparison between the observed \vsini\ and 
the computed \veq\ for the stars listed in Table~\ref{tab:variable}. As 
expected, taking into consideration the \vsini\ uncertainties, most of the 
stars fall below the equality line ($\sin i\,\leq1$). For the stars which 
fall instead above the equality line it is possible that the calculated radius 
is too large or that the measured rotational period is too small. By 
calculating the surface gravity (\logg), from \M, \Teff, and \logl, it is 
possible to identify the stars for which a problem with the stellar parameters 
is likely to be present, but this is not the case for any of our targets. 
The stars, falling above the equality line are: $\gamma$\,Ari, HD\,43819, 
HD\,47152, HD\,116114, HD\,130559, and HD\,146001. For those stars the 
calculated \logg\ values are $\leq$4.0, as expected for main sequence mCP 
stars, and therefore, we cannot exclude that their real rotational period is 
larger than that listed in Table~\ref{tab:variable}. However, at least three 
objects ($\gamma$\,Ari, HD\,47152, and HD\,130559) were found to be part of 
close visual binaries \citep{horch04,mason07}. Hence, the brightness of the
companion is included in the calculated luminosity, resulting into larger 
radii and thus smaller equatorial velocities. Since the brightness 
differences are not known, we are not able to apply the 
appropriate corrections.
%
\section{Conclusion}
We have analysed the light curves of 1028 chemically peculiar stars obtained 
with the \stereo\ spacecraft. In this work we presented the analysis and the 
results obtained for 337 magnetic chemically peculiar stars. The 
characteristics of the \stereo\ data allow the detection and study of 
photometric variations with periods between several hours and a few days and 
are therefore perfectly suitable to study rotational periods of mCP stars.

Using a matched filter algorithm we produced the light curve, phase-folded on
the best fitting period, for each star and extracted from the whole sample 
the objects which appeared clearly constant, or too badly affected by 
systematic effects to allow any reliable analysis. Those stars are listed in
Table~\ref{tab:junk}. 

For the remaining stars we performed a detailed analysis based on two different
period finding algorithms and listed the objects we classified as constant in 
Table~\ref{tab:const} and variable in Table~\ref{tab:variable}. We 
detected relevant photometric variability and measured its period for 82
mCP stars. For 48 of them this work presents the first measurement of their
rotation period, while for the remaining 34 our results are in agreement with
previous estimations. It is important to notice that the stars we classified as
constant, and therefore listed in Table~\ref{tab:junk} and \ref{tab:const},
might be intrinsically variable if their period is, for example, longer than 
10\,days or their light curve is affected by substantial blending or 
systematic effects. It is anyway likely that most of these stars are not
chemically peculair stars.

This work presents all the basic information necessary to plan detailed
spectroscopic and/or spectropolarimetric observations for 82 mCP 
stars, e.g. to perform Doppler imaging. About half of them are only suspected 
mCP stars, for which no measurements of the magnetic field have been performed. 
These stars represent a good starting point for further spectropolarimetric 
observations to improve our statistics and knowledge of the complex phenomenon 
of magnetism in stars of the upper main sequence.
\section*{Acknowledgments}
The Heliospheric Imager (HI) instrument was developed by a collaboration 
that included the Rutherford Appleton Laboratory and the University of 
Birmingham, both in the United Kingdom, and the Centre Spatial de Li\'ege 
(CSL), Belgium, and the US Naval Research Laboratory (NRL), Washington DC, 
USA. The \textit{STEREO}/SECCHI project is an international consortium 
of the Naval Research Laboratory (USA), Lockheed Martin Solar and 
Astrophysics Lab (USA), NASA Goddard Space Flight Center (USA), Rutherford 
Appleton Laboratory (UK), University of Birmingham (UK), 
Max-Planck-Institut f\"{u}r Sonnensystemforschung (Germany), Centre 
Spatial de Li\'ege (Belgium), Institut d'Optique Th\'eorique et 
Appliqu\'ee (France) and Institut d'Astrophysique Spatiale (France). 
This research has made use of the \textsc{Simbad} database, operated 
at CDS, Strasbourg, France. This research has made use of version 2.31 
\textsc{Peranso} light curve and period analysis software, maintained at 
CBA, Belgium Observatory http://www.cbabelgium.com.
Astronomy research at the Open University is supported by an STFC rolling 
grant (L.F.). KTW acknowledges support from a STFC studentship.
%

\begin{scriptsize}
\input{./tables/junkforpaper.tex}
\input{./tables/constantstarstableforpaper.tex}
\end{scriptsize}
\input{./tables/variablestableforpaper.tex}
\begin{center}
\begin{figure*}
\includegraphics[width=165mm,clip]{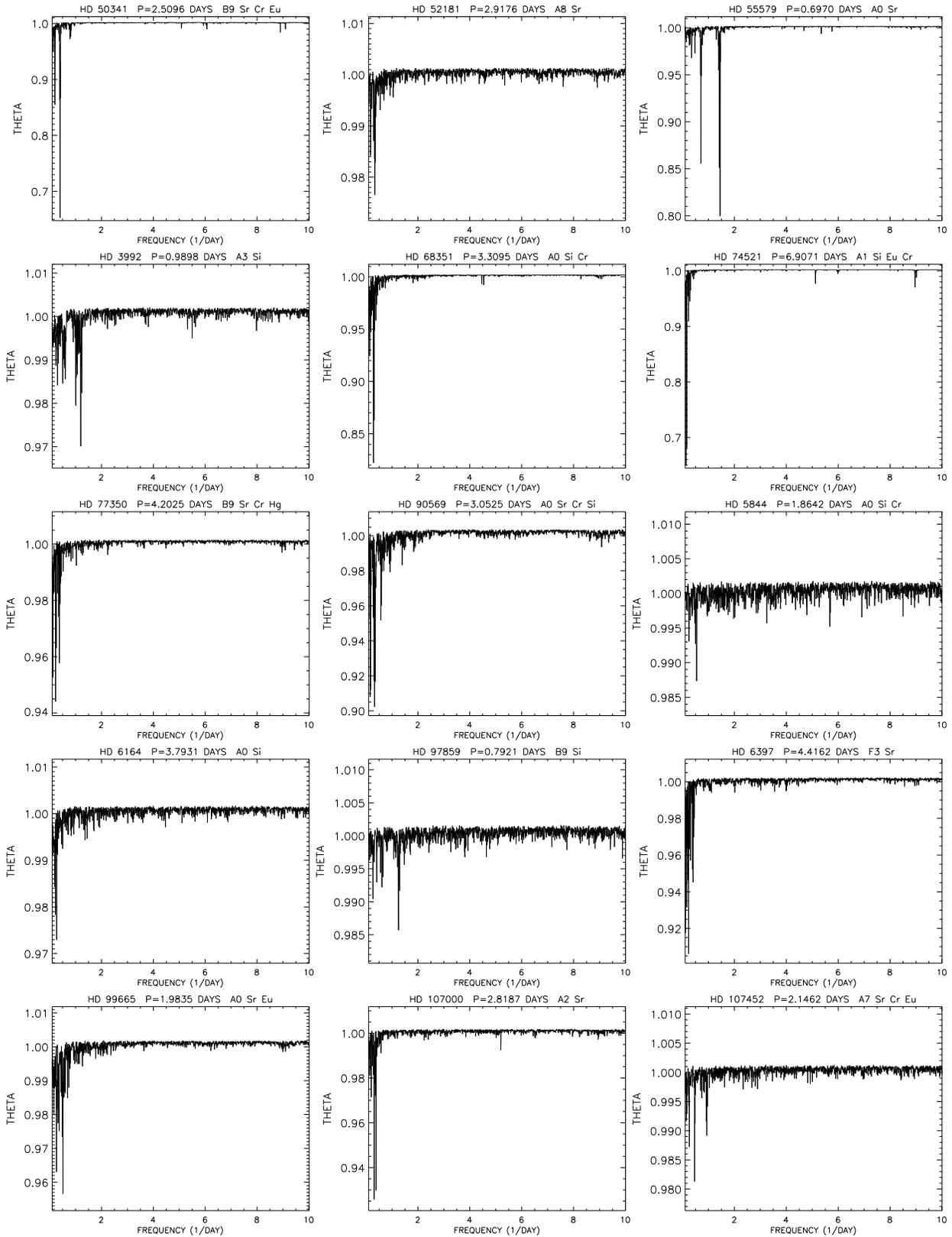}
\caption{Periodograms obtained for the mCP stars, listed in
Table~\ref{tab:variable}.}
\label{fig:allperiodograms} 
\end{figure*}
\begin{figure*}
\includegraphics[width=165mm,clip]{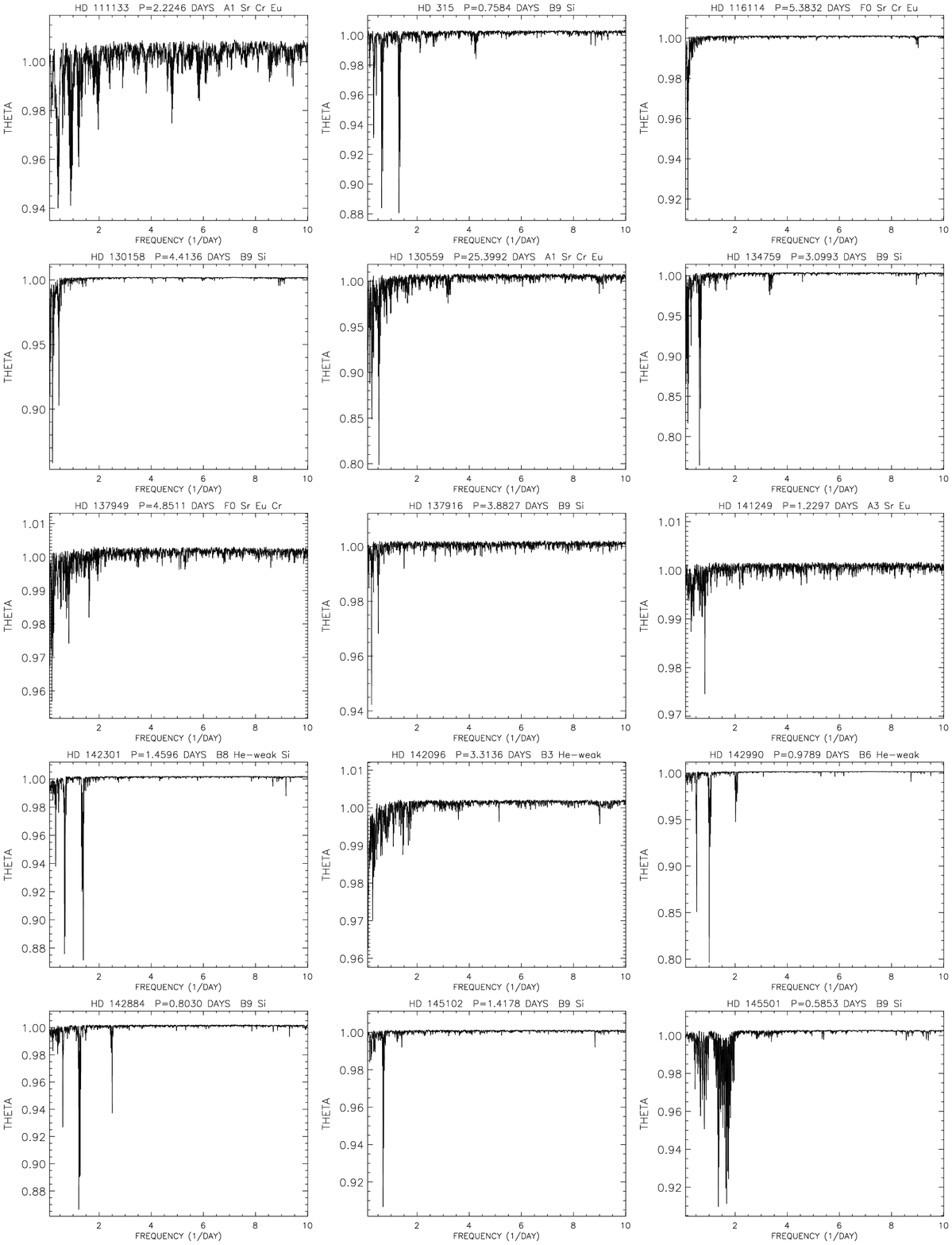}
\end{figure*}
\begin{figure*}
\includegraphics[width=165mm,clip]{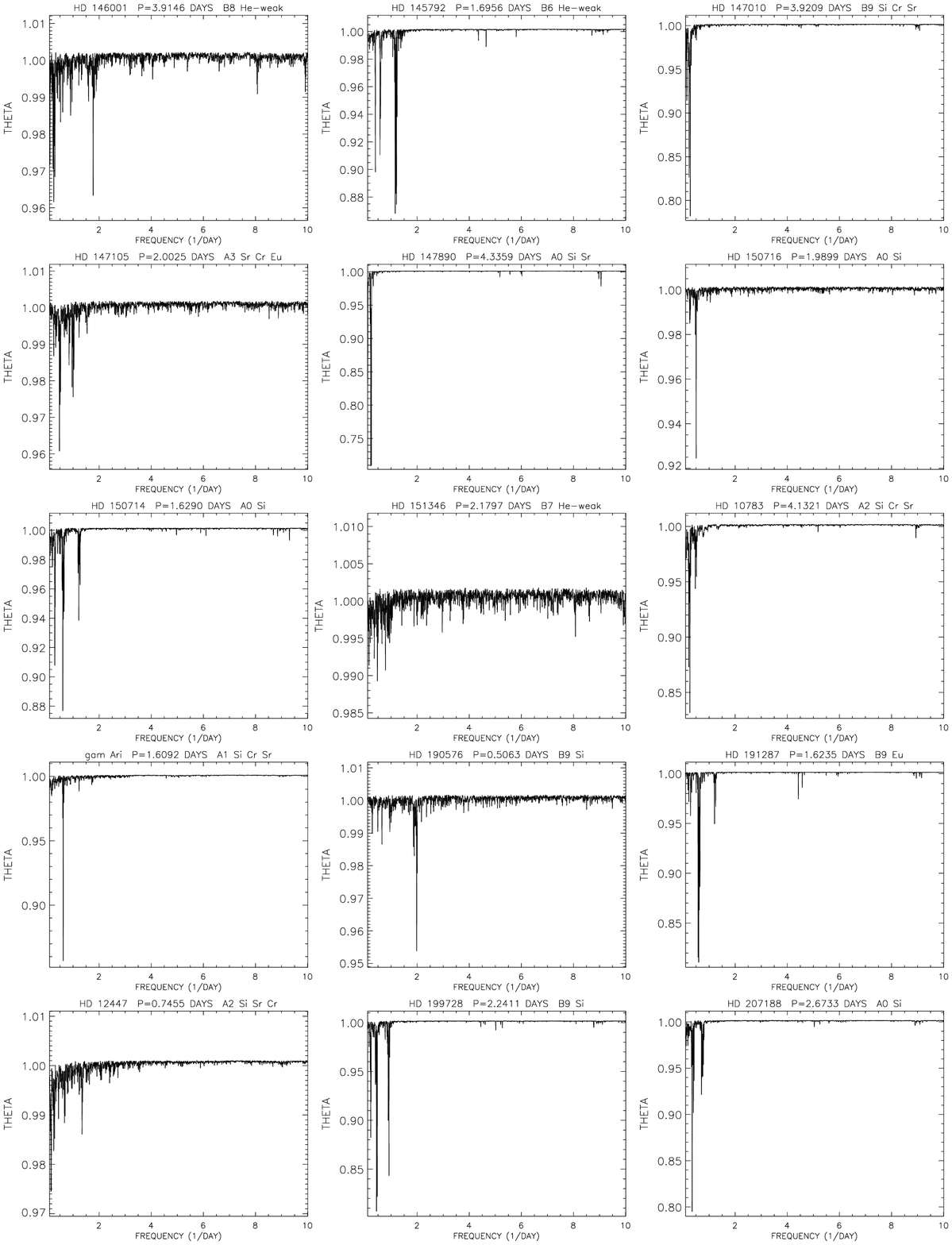}
\end{figure*}
\begin{figure*}
\includegraphics[width=165mm,clip]{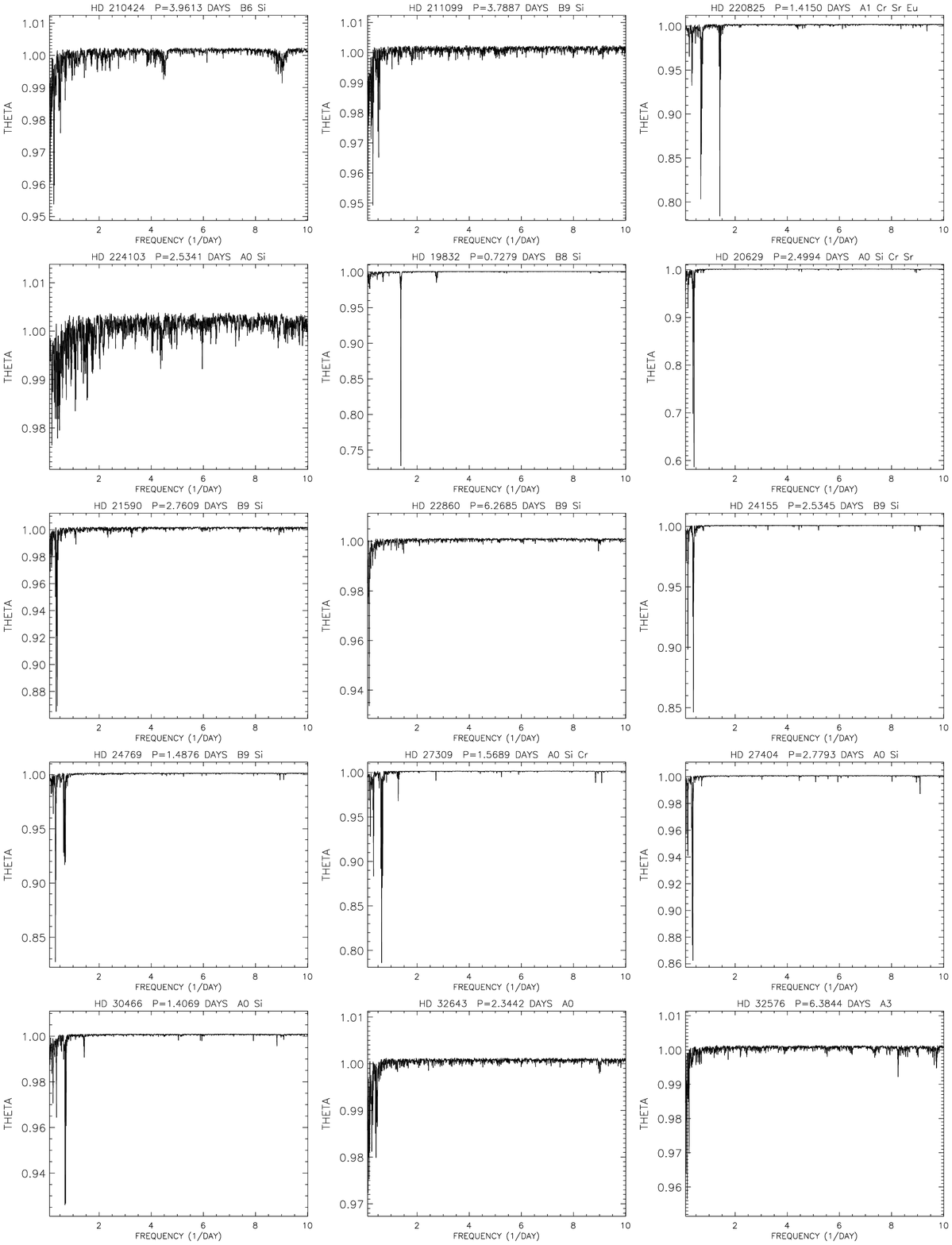}
\end{figure*}
\begin{figure*}
\includegraphics[width=165mm,clip]{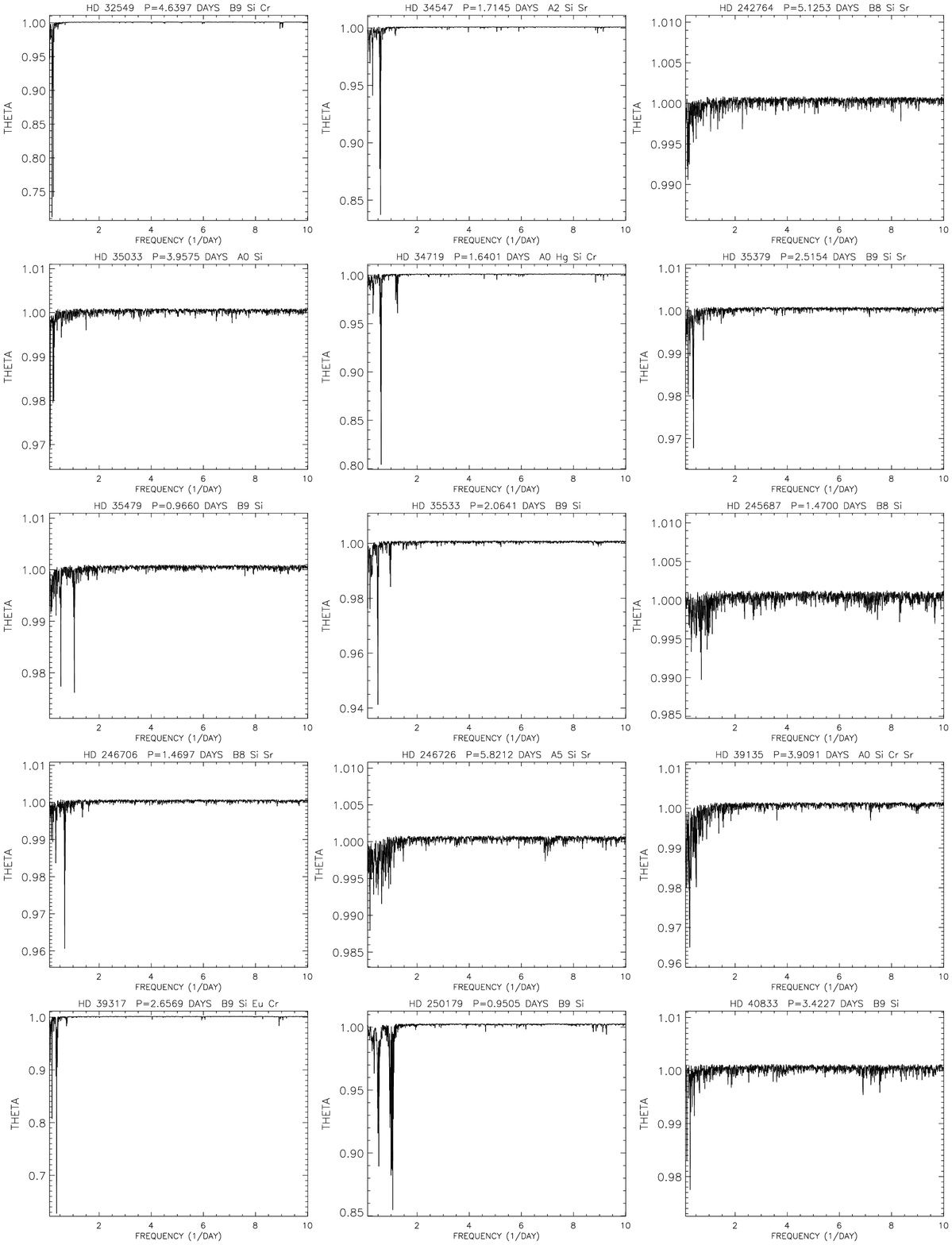}
\end{figure*}
\begin{figure*}
\includegraphics[width=165mm,clip]{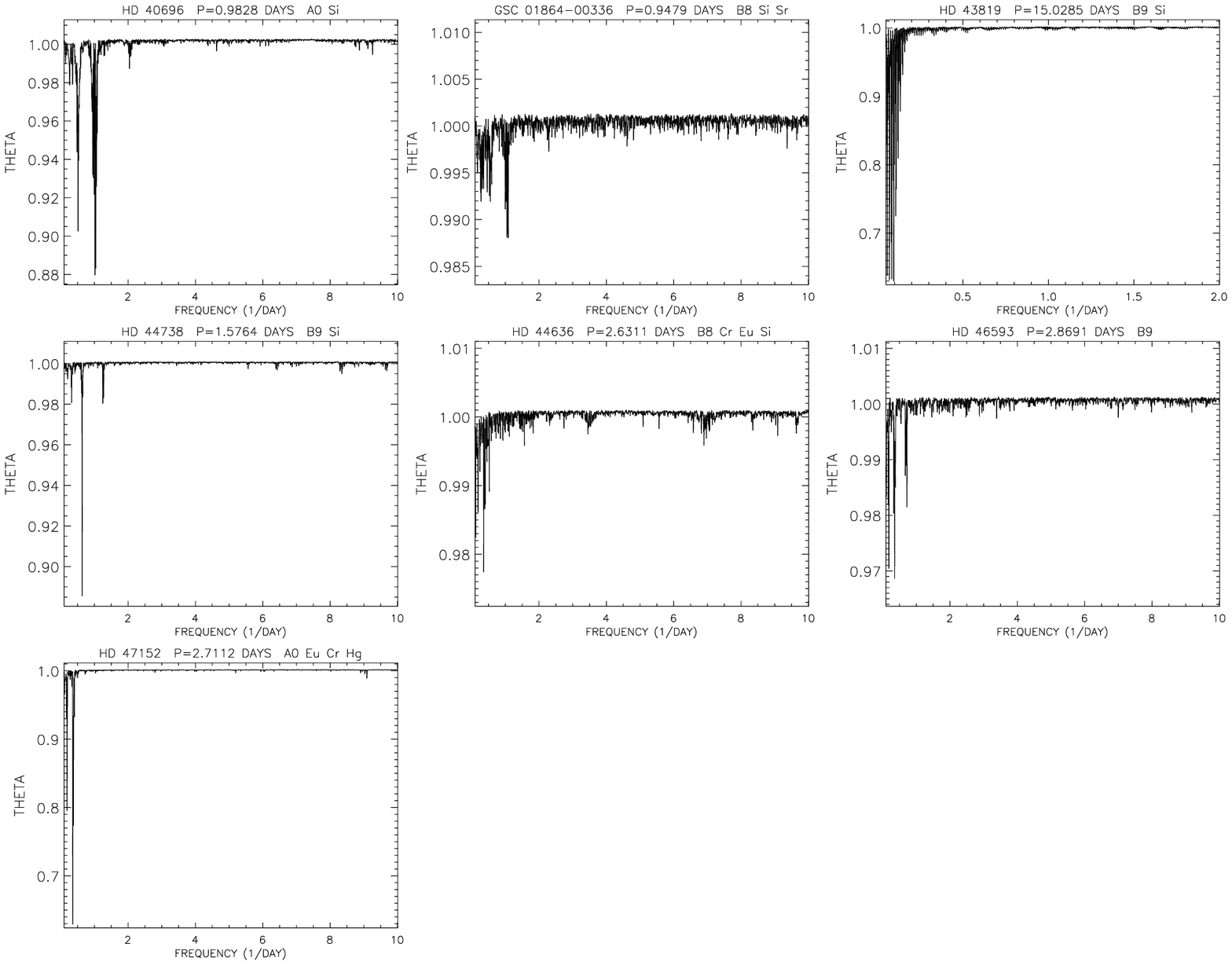}
\end{figure*}
\end{center} 
\begin{center}
\begin{figure*}
\includegraphics[width=165mm,clip]{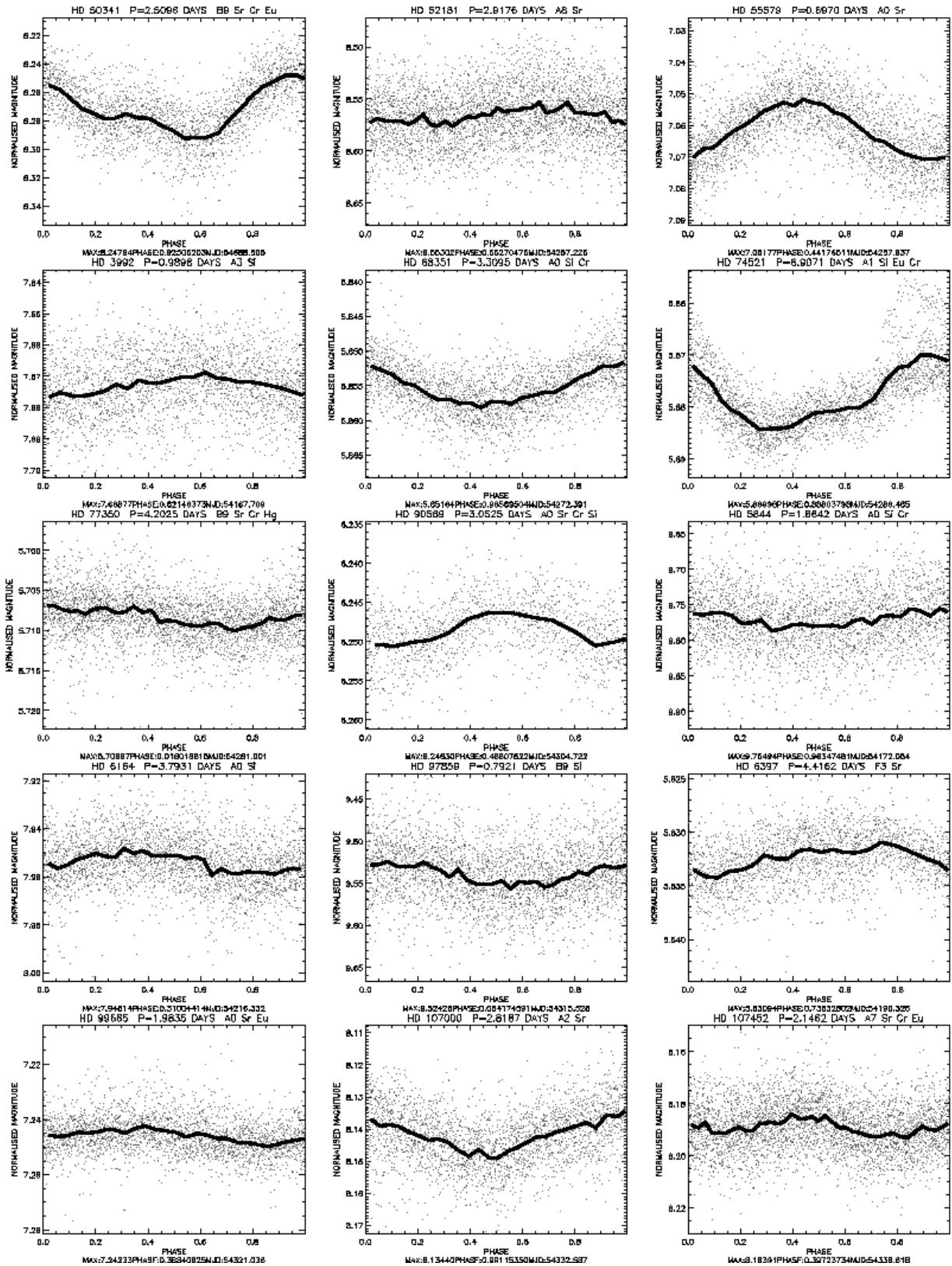}
\caption{Light curves obtained for the mCP stars, listed in
Table~\ref{tab:variable}, phase-folded on the most significant period.}
\label{fig:alllightcurves} 
\end{figure*}
\begin{figure*}
\includegraphics[width=165mm,clip]{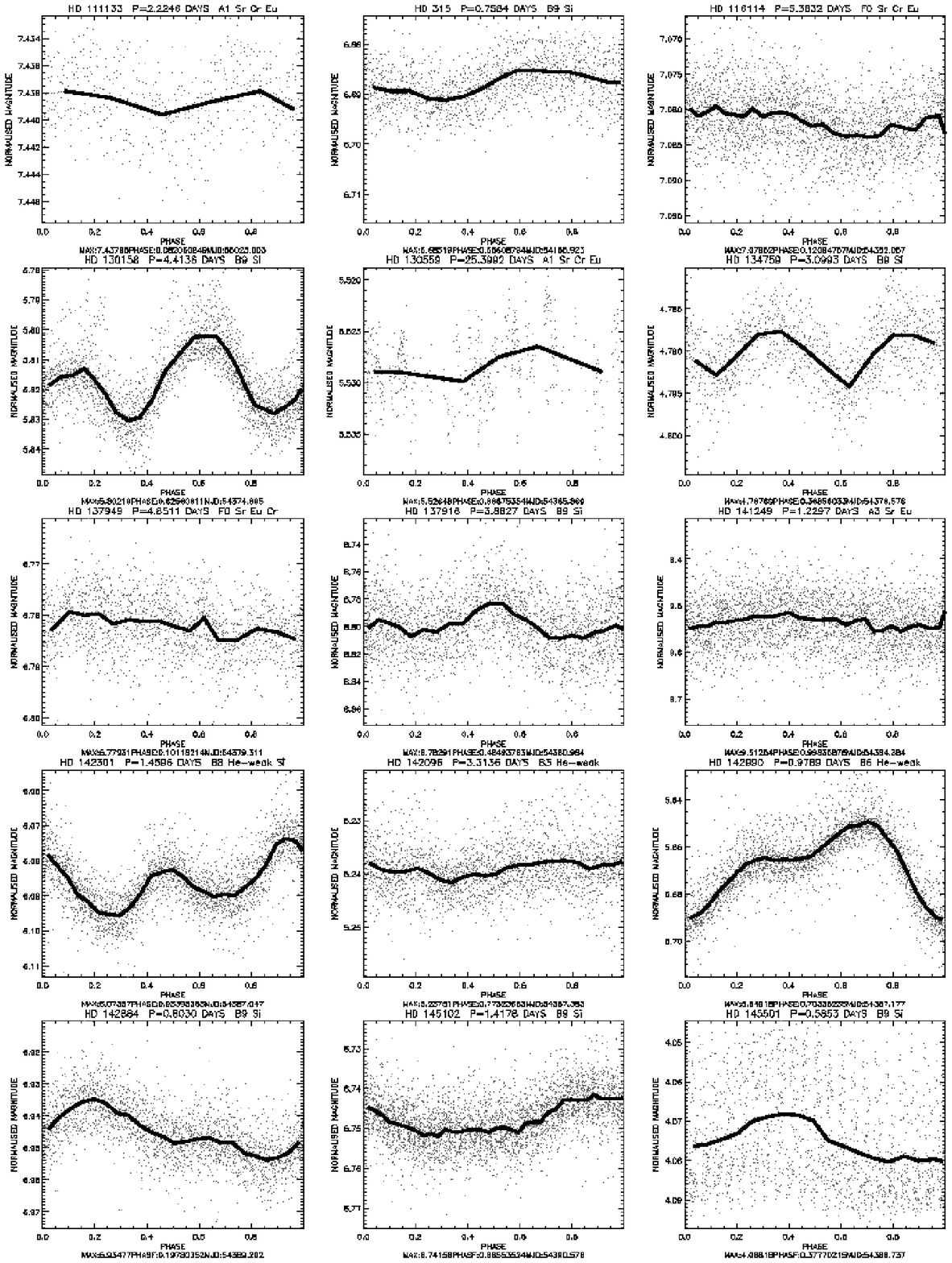}
\end{figure*}
\begin{figure*}
\includegraphics[width=165mm,clip]{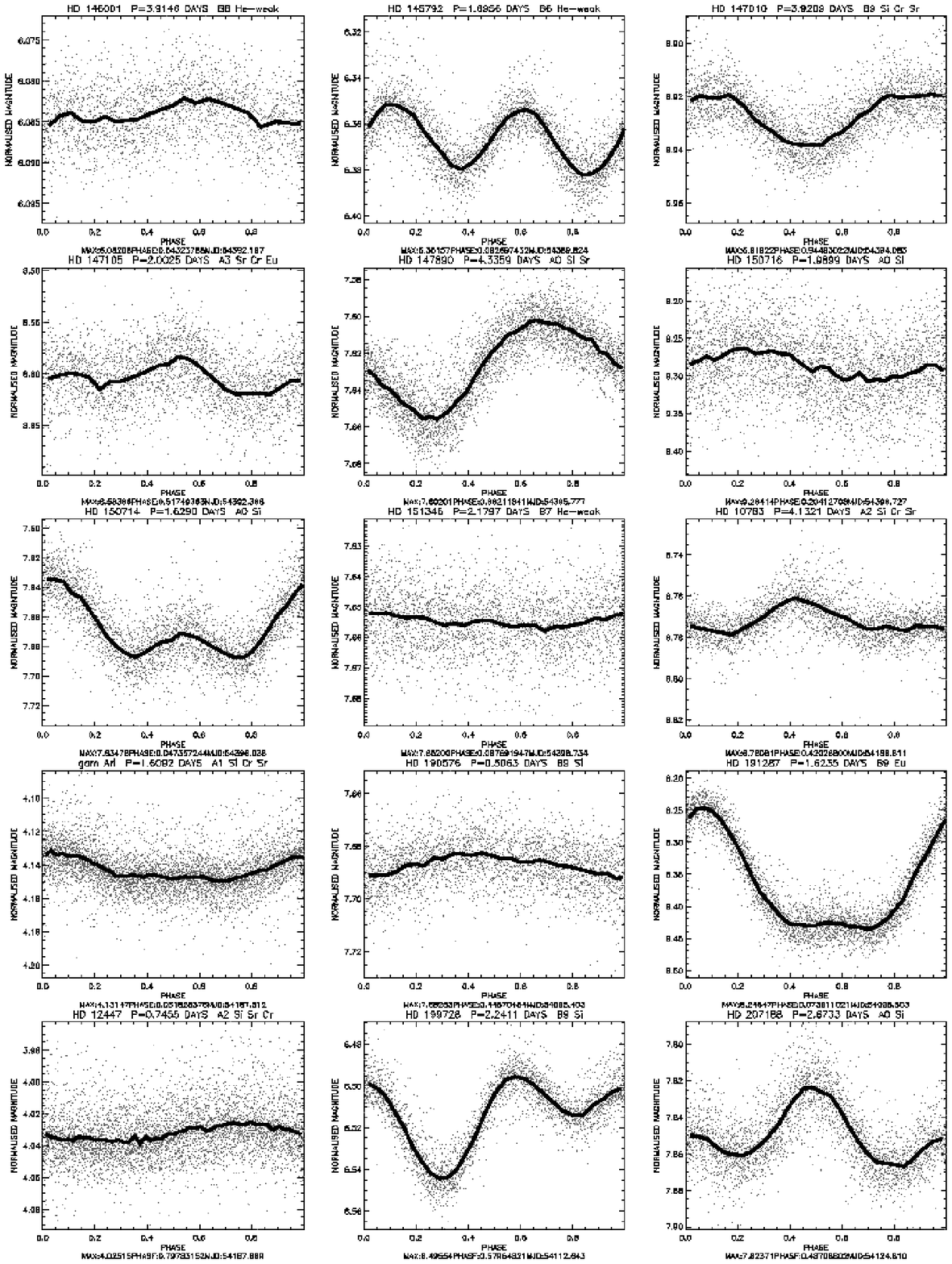}
\end{figure*}
\begin{figure*}
\includegraphics[width=165mm,clip]{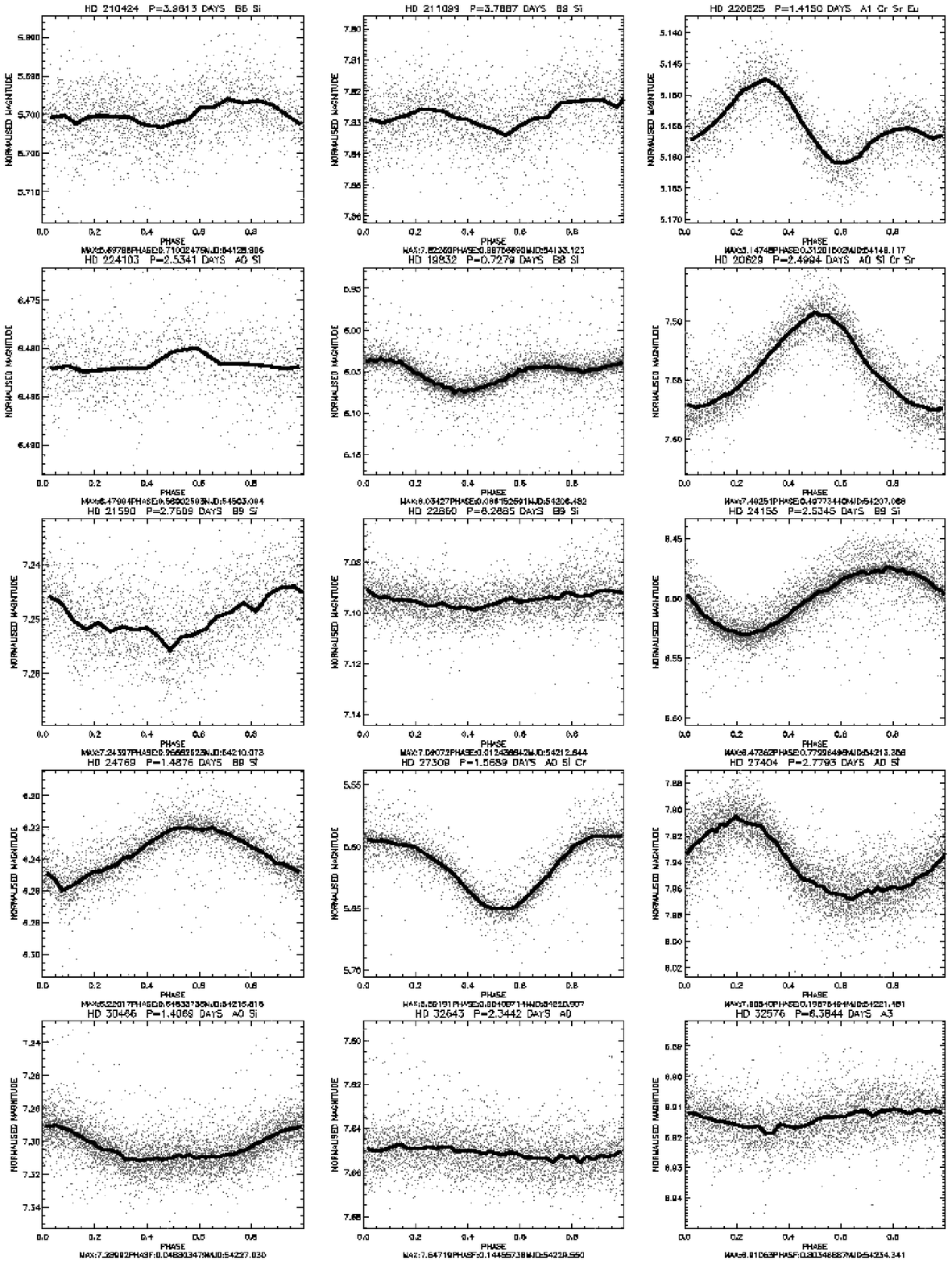}
\end{figure*}
\begin{figure*}
\includegraphics[width=165mm,clip]{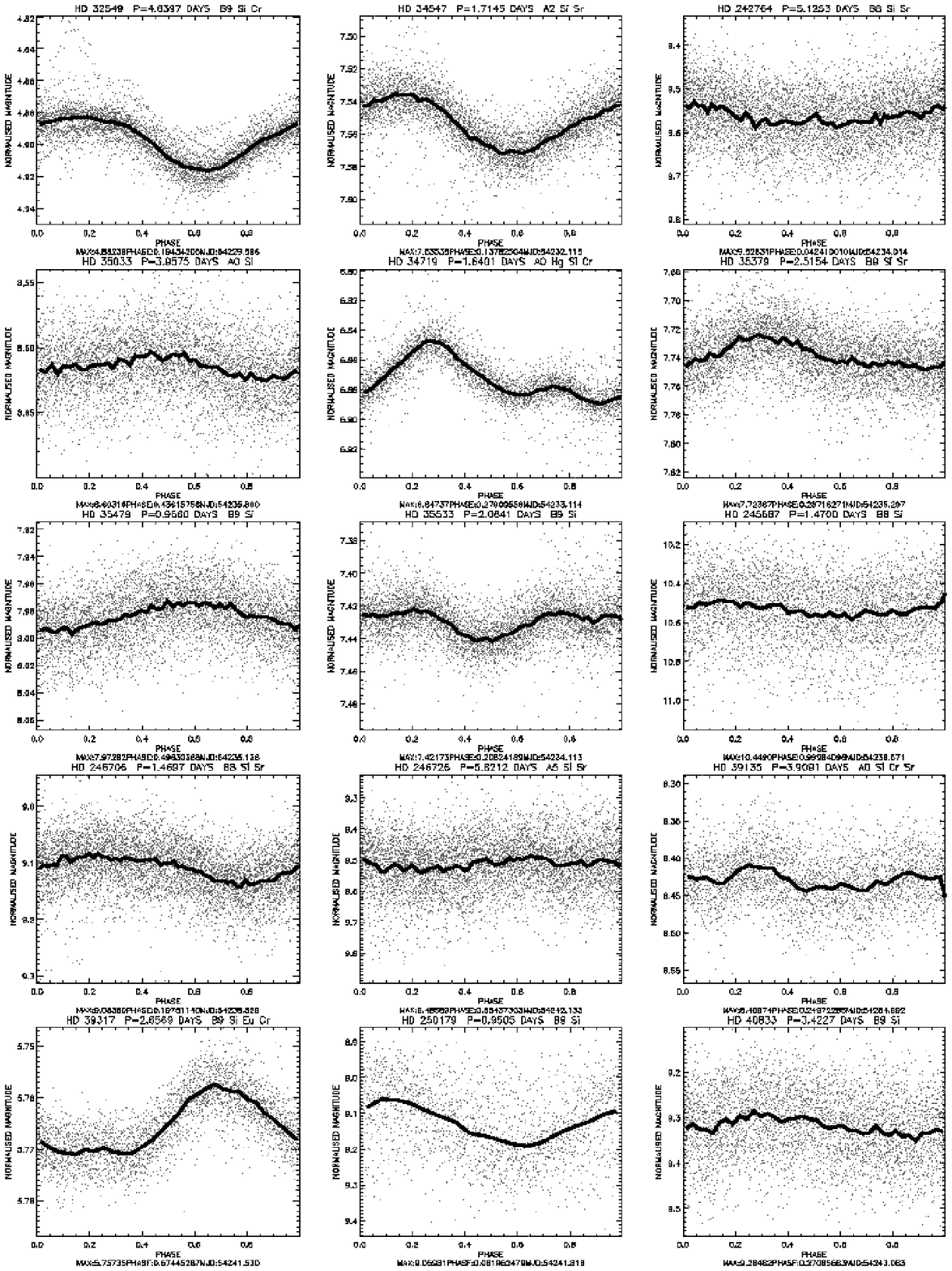}
\end{figure*}
\begin{figure*}
\includegraphics[width=165mm,clip]{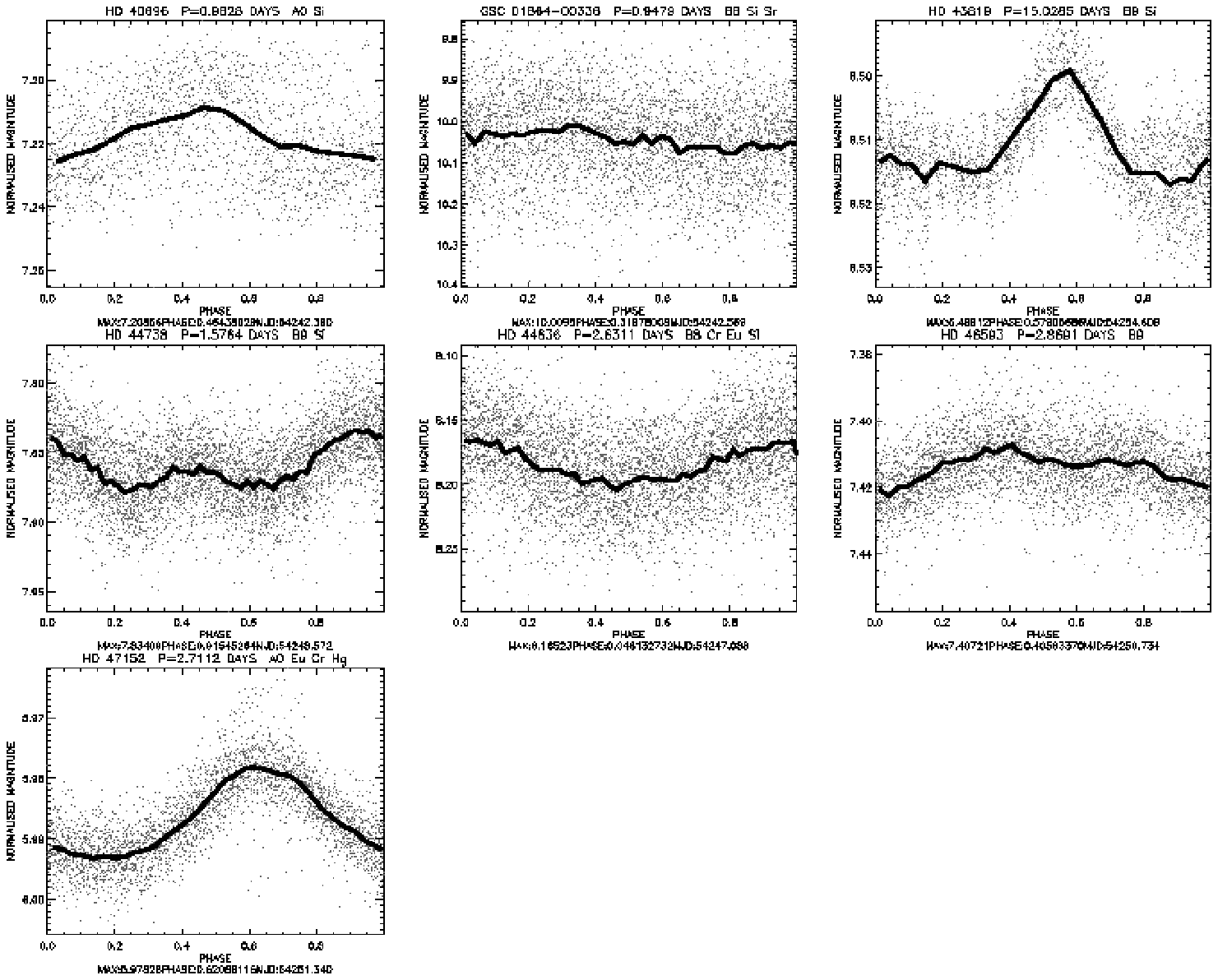}
\end{figure*}
\end{center} 

\bsp

\label{lastpage}

\end{document}

%% file: tables/junkforpaper.tex
\onecolumn
\begin{longtable}{lrrrcrcc}
\caption{Basic properties of the CP2 and CP4 stars identified as constant or for which the quality of the data
prevented the detection of any variability.}
\protect\label{tab:junk}\\
\hline\hline \\
Name & \# R09 & RA  & DEC & V  & Spectral & CP    & \vsini \\
     &        & deg.& deg.& mag& Type     & class & \kms   \\
\hline
\endfirsthead
\caption{continued.}\\
\hline\hline \\       
Name & \# R09 & RA  & DEC & V  & Spectral & CP    & \vsini \\
     &        & deg.& deg.& mag& Type     & class & \kms   \\
\hline
\endhead
\hline
\endfoot
\hline
\hline
\endlastfoot
 HD\,965	     &   160	&   3.51693 &	-0.03337 &  8.57 &   A8,Sr,Eu,Cr & CP2   & 90.0(13.5/1)  \\
 BD+09\,124	     &  1640	&  16.30230 &	10.27470 & 10.54 &	   A0,NA & CP2   &               \\
 HD\,6590	     &  1680	&  16.71320 &	15.74080 & 10.15 &	   A3,NA & CP2?  &               \\
 HD\,17471	     &  4380	&  42.19130 &	25.18810 &  5.89 &	   A0,Si & CP2   & 46.3(1.2/3)   \\
 HD\,20150	     &  5000	&  48.72540 &	21.04440 &  4.88 &	   A1,NA & CP2?  & 125.0(5.7/2)  \\
 HD\,26571	     &  6780	&  63.21350 &	22.41350 &  6.15 &	   B8,Si & CP2   & 20.0(3.0/1)   \\
 HD\,281886	     &  6900	&  64.40670 &	31.43480 &  8.90 &	   F0,Sr & CP2?  &               \\
 HD\,27295	     &  7000	&  64.85870 &	21.14230 &  5.49 &	   B9,Mn & CP2   & 9.2(0.5/6)    \\
 HD\,284639	     &  7770	&  71.49670 &	23.91680 &  9.63 &	A0,Si,Cr & CP2   &               \\
 HD\,32992	     &  8440	&  76.93260 &	14.35960 &  8.19 &	   A1,NA & CP2?  &               \\
 AAO+32\,27	     &  8844	&  80.13170 &	32.60390 & 10.79 &	   B9,Si & CP2?  &               \\
 HD\,242720	     &  8858	&  80.22580 &	30.09630 & 10.60 &	B9,Si,Sr & CP2?  &               \\
 HD\,242705	     &  8865	&  80.27430 &	32.43030 & 11.40 &	   B9,Si & CP2?  &               \\
 HD\,243010	     &  8951	&  80.70000 &	23.72580 & 10.80 &	   A3,NA & CP2?  &               \\
 HD\,243202	     &  8983	&  81.12300 &	32.81910 & 10.90 &	   B5,NA & CP4   &               \\
 HD\,243308	     &  9006	&  81.24240 &	30.07960 & 10.80 &	   B9,Si & CP2?  &               \\
 AAO+26\,15	     &  9028	&  81.29800 &	26.46530 & 10.90 &	   B9,Si & CP2?  &               \\
 BD+32\,976	     &  9024	&  81.32100 &	32.67810 & 10.27 &	   A2,Sr & CP2?  &               \\
 HD\,243321	     &  9026	&  81.32430 &	32.61640 &  9.64 &	   A0,Si & CP2?  &               \\
 HD\,243395	     &  9047	&  81.34460 &	29.36330 &  9.86 &	A0,Si,Sr & CP2?  &               \\
 HD\,243356	     &  9034	&  81.36610 &	32.66690 & 11.60 &	   A0,Si & CP2?  &               \\
 HD\,243408	     &  9052	&  81.42580 &	32.30090 & 10.10 &	   B9,Si & CP2?  &               \\
 HD\,35436	     &  9059	&  81.53600 &	32.80490 &  9.57 &	A1,Si,Sr & CP2?  &               \\
 HD\,243492	     &  9082	&  81.56870 &	33.26230 & 10.60 &	A0,Si,Sr & CP2?  &               \\
 AAO+31\,68	     &  9155	&  81.73960 &	31.06840 & 11.54 &	   B9,Sr & CP2?  &               \\
 HD\,35693	     &  9220	&  81.80790 &	15.25760 &  6.18 &	   A2,Cr & CP2?  & 77.5(8.0/2)   \\
 HD\,243791	     &  9215	&  81.95790 &	29.80130 & 11.40 &	   A0,Sr & CP2?  &               \\
 HD\,243970	     &  9246	&  82.28910 &	32.55970 & 10.07 &	B9,Si,Sr & CP2?  &               \\
 HD\,244099	     &  9266	&  82.44620 &	31.63800 & 11.00 &	   B9,Sr & CP2?  &               \\
 AAO+31\,112	     &  9337	&  82.71830 &	31.26950 & 11.95 &	   B8,Si & CP2?  &               \\
 HD\,244248	     &  9336	&  82.72660 &	32.76810 & 11.40 &	A5,Si,Sr & CP2?  &               \\
 AAO+28\,89	     &  9355	&  82.84630 &	28.23690 & 11.71 &	B9,Si,Sr & CP2?  &               \\
 HD\,244352	     &  9351	&  82.85880 &	30.69890 & 10.70 &	   B9,Si & CP2?  &               \\
 HD\,244876	     &  9553	&  83.62420 &	29.26920 & 10.23 &	A0,Si,Sr & CP2?  &               \\
 AAO+27\,87	     &  9572	&  83.65970 &	27.23250 & 10.70 &	B9,Si,Sr & CP2?  &               \\
 HD\,244933	     &  9576	&  83.73770 &	32.56150 & 11.50 &	   A1,Si & CP2?  &               \\
 HD\,244955	     &  9613	&  83.76450 &	30.18470 & 12.10 &	    A,Sr & CP2?  &               \\
 HD\,245112	     &  9656	&  83.87710 &	23.28400 & 10.90 &	A0,Si,Cr & CP2?  &               \\
 HD\,245155	     &  9662	&  83.92830 &	25.27480 &  9.86 &	B9,Si,Sr & CP2?  &               \\
 HD\,245153	     &  9664	&  83.96980 &	28.16360 & 10.33 &	A0,Si,Sr & CP2?  &               \\
 HD\,245191	     &  9698	&  84.07970 &	30.27230 & 11.20 &	   A0,Si & CP2?  &               \\
 HD\,245222	     &  9748	&  84.09270 &	27.90150 & 10.80 &	A5,Cr,Si & CP2?  &               \\
 HD\,245320	     &  9837	&  84.21000 &	28.19000 & 10.56 &   B9,Si,Cr,Sr & CP2?  &               \\
 HD\,245353	     &  9858	&  84.26490 &	27.62960 & 10.22 &	A1,Si,Sr & CP2?  &               \\
 HD\,245416	     &  9876	&  84.38420 &	31.77830 &  9.43 &	A2,Cr,Si & CP2?  &               \\
 HD\,245725	     & 10002	&  84.77580 &	32.63670 & 11.00 &	   A0,Si & CP2?  &               \\
 HD\,245726	     & 10008	&  84.78270 &	32.22540 & 10.80 &	A1,Si,Sr & CP2?  &               \\
 HD\,245786	     & 10012	&  84.80020 &	29.05760 & 11.20 &	    A,Si & CP2?  &               \\
 HD\,245990	     & 10072	&  84.99170 &	26.37620 & 10.70 &	   A1,Si & CP2?  &               \\
 AAO+32\,281	     & 10047	&  85.00080 &	32.90320 & 11.22 &	B9,Si,Sr & CP2?  &               \\
 HD\,246148	     & 10121	&  85.18930 &	23.52640 & 10.60 &	   B9,Si & CP2?  &               \\
 AAO+33\,312	     & 10107	&  85.25890 &	33.39200 & 11.30 &	   B9,Si & CP2?  &               \\
 HD\,246276	     & 10157	&  85.37980 &	25.99710 & 10.80 &	   B9,Si & CP2?  &               \\
 HD\,246587	     & 10224	&  85.71060 &	24.24100 & 11.00 &   A2,Si,Cr,Sr & CP2?  &               \\
 HD\,246685	     & 10231	&  85.86690 &	27.47300 & 11.70 &	 A,Si,Sr & CP2?  &               \\
 HD\,246861	     & 10243	&  86.13330 &	28.53510 &  9.94 &	   B9,Si & CP2?  &               \\
 HD\,246993	     & 10258	&  86.21710 &	25.29510 & 10.80 &	   B9,Si & CP2?  &               \\
 HD\,246970	     & 10254	&  86.21970 &	28.56570 & 10.60 &	   B9,Sr & CP2?  &               \\
 AAO+30\,227	     & 10314	&  86.54500 &	30.92590 & 12.11 &	B9,Si,Sr & CP2?  &               \\
 AAO+27\,175	     & 10323	&  86.62690 &	27.58840 & 11.63 &	   B8,Si & CP2?  &               \\
 AAO+29\,209	     & 10329	&  86.71300 &	29.56960 & 10.77 &     B8, Si,Cr & CP2?  &               \\
 HD\,247437	     & 10332	&  86.79040 &	31.90220 & 10.04 &	B9,Si,Sr & CP2?  &               \\
 HD\,247591	     & 10368	&  86.98250 &	29.49110 & 10.30 &   A1,Si,Cr,Sr & CP2?  &               \\
 HD\,247629	     & 10379	&  87.09280 &	33.44420 & 10.80 &	   A0,Sr & CP2?  &               \\
 AAO+33\,402	     & 10381	&  87.09350 &	33.22400 & 11.92 &	B9,Si,Sr & CP2?  &               \\
 HD\,247794	     & 10412	&  87.29340 &	32.47900 & 10.70 &	   A0,Sr & CP2?  &               \\
 HD\,247833	     & 10422	&  87.35370 &	32.83290 & 10.80 &	   B9,Sr & CP2?  &               \\
 HD\,274959	     & 10433	&  87.41930 &	26.94220 & 10.30 &	   B9,Sr & CP2?  &               \\
 HD\,247931	     & 10427	&  87.46170 &	33.45890 & 10.30 &	   B7,Si & CP2?  &               \\
 HD\,248131	     & 10463	&  87.69020 &	31.94080 & 10.60 &	B9,Si,Sr & CP2?  &               \\
 AAO+31\,309	     & 10498	&  87.89430 &	31.74490 & 11.30 &	   A2,Si & CP2?  &               \\
 AAO+32\,406	     & 10509	&  88.00800 &	32.20760 & 11.62 &	   A0,Sr & CP2?  &               \\
 HD\,39200	     & 10513	&  88.04520 &	29.91650 &  9.33 &	   B9,Si & CP2?  &               \\
 HD\,248582	     & 10551	&  88.14810 &	24.22680 &  9.75 &	   A0,Si & CP2?  &               \\
 HD\,248619	     & 10561	&  88.26190 &	28.55520 & 10.67 &	B9,Si,Sr & CP2?  &               \\
 HD\,248727	     & 10588	&  88.44110 &	33.28710 & 10.20 &   A0,Mn,Si,Cr & CP2?  &               \\
 HD\,248891	     & 10603	&  88.62590 &	31.16230 & 10.05 &	   A0,Si & CP2?  &               \\
 HD\,248943	     & 10607	&  88.63820 &	28.90100 & 10.60 &	   A0,Si & CP2?  &               \\
 HD\,249121	     & 10632	&  88.86340 &	24.03890 & 11.17 &	   A1,Sr & CP2?  &               \\
 HD\,249279	     & 10639	&  89.08810 &	28.73570 &  9.86 &	A0,Si,Sr & CP2?  &               \\
 HD\,249401	     & 10663	&  89.25850 &	29.54600 & 10.80 &	   A2,Si & CP2?  &               \\
 HD\,249478	     & 10668	&  89.35870 &	29.88800 & 10.44 &	A1,Si,Sr & CP2?  &               \\
 HD\,40038	     & 10669	&  89.42110 &	28.29090 &  8.55 &	   A0,Sr & CP2?  &               \\
 HD\,249664	     & 10697	&  89.56630 &	29.14840 & 10.65 &	A1,Si,Sr & CP2?  &               \\
 AAO+28\,373	     & 10718	&  89.62280 &	28.73300 & 11.42 &	B8,Si,Sr & CP2?  &               \\
 HD\,249931	     & 10793	&  89.95430 &	30.94990 & 10.60 &	A0,Si,Sr & CP2?  &               \\
 AAO+28\,401	     & 10813	&  90.10110 &	28.70450 & 11.98 &	   A0,Si & CP2?  &               \\
 HD\,250115	     & 10819	&  90.15560 &	30.08410 & 10.14 &	   B8,Si & CP2?  &               \\
 HD\,40678	     & 10876	&  90.37170 &	23.70390 &  7.35 &	A0,Si,Sr & CP2   &               \\
 HD\,250515	     & 10915	&  90.61390 &	28.49550 & 10.21 &	   A0,Si & CP2?  &               \\
 HD\,250662	     & 10931	&  90.71950 &	25.67540 &  8.96 &	B9,Si,Sr & CP2?  &               \\
 AAO+28\,455	     & 10957	&  90.90330 &	28.53770 & 12.04 &	   B8,Si & CP2?  &               \\
 AAO+27\,400	     & 10965	&  90.94320 &	27.71830 & 11.28 &	B9,Si,Sr & CP2?  &               \\
 HD\,250827	     & 10970	&  91.02330 &	32.88760 &  9.37 &	   A3,Sr & CP2?  &               \\
 HD\,250962	     & 11003	&  91.13700 &	30.23840 &  9.42 &	   B9,Si & CP2?  &               \\
 BD+25\,1117	     & 11094	&  91.60310 &	25.83430 & 10.80 &	   B8,Si & CP2?  &               \\
 HD\,252076	     & 11157	&  92.09630 &	23.92020 & 10.02 &	B8,Si,Sr & CP2?  &               \\
 HD\,252286	     & 11186	&  92.28060 &	24.98930 & 10.20 &	   A0,Sr & CP2?  &               \\
 AAO+28\,567	     & 11196	&  92.45440 &	28.82590 & 11.72 &	   B9,Sr & CP2?  &               \\
 HD\,252398	     & 11194	&  92.50550 &	32.88770 &  9.67 &	B9,Cr,Si & CP2?  &               \\
 HD\,42477	     & 11310	&  92.86630 &	13.63860 &  6.04 &	   A1,Si & CP2   & 220.0(33.0/1) \\
 HD\,42509	     & 11320	&  93.00560 &	19.79050 &  5.75 &	   B9,Si & CP2?  & 92.5(9.0/2)   \\
 HD\,259380	     & 12410	&  98.49450 &	31.91610 & 10.62 &	   A7,Sr & CP2?  &               \\
 BD+21\,1501	     & 14670	& 106.32400 &	21.03770 & 10.30 &	   NA,Si & CP2   &               \\
 HD\,68542	     & 18990	& 123.47300 &	27.77860 &  7.77 &	F3-F7,Sr & CP2?  & 27.0(4.1/1)   \\
 HD\,72943	     & 20230	& 129.03200 &	15.31360 &  6.33 &	 A5-F,NA & CP2?  & 63.7(1.4/4)   \\
 BD-02\,3229	     & 27140	& 162.39799 &	-3.38271 &  9.78 &	   NA,Sr & CP2?  &               \\
 HD\,96097	     & 27670	& 166.25400 &	 7.33601 &  4.63 &	   F3,Sr & CP2?  & 25.2(1.3/5)   \\
 HD\,98664	     & 28450	& 170.28400 &	 6.02933 &  4.04 &	   B9,Si & CP2?  & 60.6(0.4/5)   \\
 HD\,104833	     & 30360	& 181.07700 &	-9.35832 &  9.04 &	   F0,Sr & CP2   & 40.4(1.7/1)   \\
 HD\,109704	     & 31810	& 189.19701 &	-5.83190 &  5.88 &	   A2,NA & CP2   & 140.0(8.0/1)  \\
 HD\,111702	     & 32410	& 192.83299 &  -11.84510 &  8.81 &	   F0,Sr & CP2   & 54.9(6.0/1)   \\
 HD\,123255	     & 35336	& 211.67799 &	-9.31351 &  5.46 &	   F1,Cr & CP2?  & 154.7(9.4/5)  \\
 HD\,134214	     & 38100	& 227.25999 &  -13.99960 &  7.48 &   F2,Sr,Eu,Cr & CP2   &               \\
 HD\,142250	     & 40360	& 238.62500 &  -27.33860 &  6.10 &    B6,He-weak & CP4   & 34.0(1.3/3)   \\
 HD\,142502	     & 40430	& 238.85800 &  -15.04090 &  9.50 &	A5,Sr,Eu & CP2   &               \\
 HD\,144941	     & 41070	& 242.35201 &  -27.22730 & 10.02 &	   B8,He & CP4?  &               \\
 HD\,146607	     & 41420	& 244.56700 &  -30.95540 & 10.18 &	   A0,Si & CP2   &               \\
 HD\,148117	     & 41860	& 246.75800 &  -26.74220 & 10.70 &   A7,Si,Eu,Cr & CP2   &               \\
 HD\,149151	     & 42180	& 248.54601 &  -31.28480 &  8.12 &	   A0,Si & CP2   &               \\
 HD\,150347	     & 42510	& 250.39799 &  -28.58640 &  8.97 &	   B9,Si & CP2   &               \\
 HD\,151560	     & 42870	& 252.31400 &  -28.71010 &  9.73 &	   A3,Sr & CP2   &               \\
 HD\,192723	     & 53790	& 304.34100 &  -28.49830 & 10.41 &	   A0,Si & CP2   &               \\
 HD\,196691	     & 54900	& 309.79400 &	-6.15760 &  8.63 &	   A0,Si & CP2?  &               \\
 BD-12\,5801	     & 54930	& 310.07199 &  -12.23580 &  9.70 &	   NA,Sr & CP2   &               \\
 HD\,198216	     & 55190	& 312.24700 &	-5.74948 &  9.94 &	   A0,NA & CP2?  &               \\
 HD\,202406	     & 56400	& 318.97900 &	-9.39114 &  7.85 &	   F2,Sr & CP2   &               \\
 HD\,207052	     & 57620	& 326.63400 &  -11.36600 &  5.57 &	   A1,NA & CP2   & 192.5(5.7/2)  \\
 HD\,209051	     & 58130	& 330.16699 &	-6.43251 &  8.76 &   A0,Si,Cr,Eu & CP2   &               \\
 HD\,215624	     & 59570	& 341.67200 &  -13.55160 & 10.56 &	A2,Sr,Si & CP2?  &               \\
 HD\,216018	     & 59680	& 342.35999 &  -11.34920 &  7.61 &   A7,Sr,Eu,Cr & CP2   &               \\
 HD\,219831	     & 60350	& 349.75601 &	-9.05975 & 10.10 &	   A2,Sr & CP2   &               \\
 HD\,222962	     & 61143	& 356.34100 &	10.17980 &  6.55 &	   A4,NA & CP2?  &               \\
\hline                                                  
\end{longtable}

%% file: tables/constantstarstableforpaper.tex
\onecolumn
\begin{longtable}{lrrrcrcccrrc}
\caption{Basic properties of the CP2 and CP4 stars identified as constant or probably constant after the individualised analysis. 
The relevant references are given at the end of Table~\ref{tab:variable}.}
\protect\label{tab:const}\\
\hline\hline \\
Name & \# R09 & RA  & DEC & V  & Spectral & CP    & Period & $\sigma_{period}$ & Analysis & Literature & \vsini \\
     &        & deg.& deg.& mag& Type     & class & days   & days              & remarks  & period     & \kms	\\
\hline
\endfirsthead
\caption{continued.}\\
\hline\hline \\       
Name & \# R09 & RA  & DEC & V  & Spectral & CP    & Period & $\sigma_{period}$ & Analysis & Literature & \vsini \\
     &        & deg.& deg.& mag& Type     & class & days   & days              & remarks  & period     & \kms	\\
\hline
\endhead
\hline
\endfoot
\hline
\hline
\endlastfoot
 HD\,224926	     & 61680	&   0.45603 &	-3.02750 &  5.12 & B8,He-weak,Mn & CP4   &   3.1604 &	0.0036 & BW   & 	    & 67.5(6.4/2)  \\
 HD\,1758	     & 402	&   5.45132 &	 0.91719 &  9.13 &	   A2,Sr & CP2   &   1.4482 &	0.0016 & W    & 	    &               \\
 HD\,10809	     & 2680	&  26.47920 &	 3.41756 &  6.82 &   A8,Sr,Cr,Eu & CP2   &	    &	       & C    & K07:2.693   &               \\
 HD\,22374	     & 5660	&  54.24180 &	23.21110 &  6.73 &   A1,Cr,Sr,Si & CP2   &	    &	       & B    & C98:10.6    & 5.0(0.8/1)    \\
 HD\,23387	     & 5980	&  56.40740 &	24.33560 &  7.19 &	A0,Cr,Si & CP2   &	    &	       & B    & 	    & 23.5(2.1/2)   \\
 HD\,23850	     & 6100	&  57.29060 &	24.05340 &  3.62 &    B8,He-weak & CP4   &   2.4624 &	0.0016 & BW   & 	    & 182.5(18.0/2) \\
 HD\,23964	     & 6140	&  57.49190 &	23.84870 &  6.74 &   B9,Si,Sr,Cr & CP2   &   1.5814 &	0.0008 & BW   & 	    & 39.0(2.8/2)   \\
 HD\,27778	     & 7120	&  65.99070 &	24.30390 &  6.33 &	   A1,Si & CP2   &	    &	       & B    & 	    &               \\
 GSC\,02390-00208    & 8795	&  79.75760 &	30.10350 & 10.80 &	A2,Si,Sr & CP2?  &   1.8444 &	0.0019 & W    & 	    &               \\
 HD\,242692	     & 8847	&  80.22540 &	33.08600 &  9.65 &	   B9,Si & CP2   &   0.6180 &	0.0001 & BW   & 	    &               \\
 HD\,242800	     & 8891	&  80.40920 &	32.75880 & 10.30 &	   A0,Si & CP2?  &	    &	       & B    & 	    &               \\
 HD\,243007	     & 8954	&  80.78520 &	30.22440 & 10.20 &	   A1,Si & CP2?  &	    &	       & C    & 	    &               \\
 GSC\,02394-00537    & 8947	&  80.78660 &	32.52450 & 10.90 &	   B9,Si & CP2?  &	    &	       & C    & 	    &               \\
 AAO+27\,25	     & 9058	&  81.40010 &	27.07680 & 11.61 &	A0,Si,Sr & CP2?  &   2.4922 &	0.0009 & W    & 	    &               \\
 HD\,243378	     & 9048	&  81.40560 &	33.26270 & 10.90 &	A0,Si,Sr & CP2?  &	    &	       & C    & 	    &               \\
 AAO+33\,110	     & 9051	&  81.42540 &	33.31920 & 11.33 &	B9,Si,Sr & CP2?  &	    &	       & C    & 	    &               \\
 HD\,243494	     & 9080	&  81.55130 &	32.03430 &  9.68 &	   B9,Si & CP2   &	    &	       & B    & 	    &               \\
 GSC\,02403-00597    & 9104	&  81.58790 &	31.21740 & 11.40 &   A0,Si,Cr,Sr & CP2?  &   2.7865 &	0.0029 & BW   & 	    &               \\
 AAO+30\,76	     & 9132	&  81.62010 &	30.36130 & 11.33 &	   A0,Sr & CP2   &	    &	       & B    & 	    &               \\
 HD\,243523	     & 9106	&  81.62030 &	33.30000 & 11.10 &	A0,Si,Sr & CP2   &	    &	       & B    & 	    &               \\
 HD\,243954	     & 9240	&  82.21120 &	29.04410 &  9.97 &	   A1,Si & CP2   &	    &	       & C    & 	    &               \\
 HD\,244303	     & 9345	&  82.80410 &	32.71930 & 11.40 &	    A,Sr & CP2?  &	    &	       & C    & 	    &               \\
 AAO+32\,173	     & 9348	&  82.83030 &	32.61440 & 11.89 &	   B9,Si & CP2?  &	    &	       & C    & 	    &               \\
 HD\,244372	     & 9354	&  82.88850 &	31.63940 & 10.60 &	   A1,Sr & CP2?  &	    &	       & C    & 	    &               \\
 HD\,244391	     & 9357	&  82.92280 &	31.62360 & 10.40 &	B8,Si,Sr & CP2?  &	    &	       & C    & 	    &               \\
 HD\,244531	     & 9376	&  83.14980 &	33.24090 & 11.45 &	   B9,Si & CP2?  &	    &	       & C    & 	    &               \\
 HD\,244737	     & 9485	&  83.48850 &	32.99670 & 11.00 &	   A0,Si & CP2?  &	    &	       & C    & 	    &               \\
 BD+26\,859	     & 9626	&  83.78830 &	26.22660 & 10.70 &	B8,Si,Sr & CP2?  &	    &	       & C    & 	    &               \\
 BD+23\,959	     & 9792	&  84.09180 &	23.60270 & 10.25 &	   A0,Si & CP2?  &	    &	       & C    & 	    &               \\
 AAO+28\,129	     & 9833	&  84.19650 &	28.18920 & 11.59 &	   B9,Sr & CP2?  &	    &	       & C    & 	    &               \\
 HD\,37098	     & 9865	&  84.28690 &	26.92430 &  5.83 &	   B9,Si & CP2   &	    &	       & B    & 	    & 50.0(7.5/1)   \\
 HD\,245423	     & 9872	&  84.31840 &	27.26770 & 10.55 &	   A3,Si & CP2?  &	    &	       & C    & 	    &               \\
 GSC\,01873-00586    & 9953	&  84.53710 &	29.01000 &  9.86 &	B9,Si,Sr & CP2?  &	    &	       & C    & 	    &               \\
 HD\,245839	     & 10033	&  84.85310 &	28.65370 & 10.90 &	   A1,Si & CP2?  &	    &	       & C    & 	    &               \\
 GSC\,01873-00024    & 10108	&  85.19830 &	28.77060 & 11.00 &	   B9,Si & CP2?  &	    &	       & C    & 	    &               \\
 HD\,246380	     & 10178	&  85.54000 &	29.25040 &  9.50 &	B8,Si,Sr & CP2?  &	    &	       & C    & 	    &               \\
 GSC\,01870-00880    & 10233	&  85.88120 &	27.17130 & 10.70 &	   A1,Si & CP2?  &	    &	       & C    & 	    &               \\
 HD\,246820	     & 10239	&  86.06830 &	27.95520 & 10.70 &	A0,Si,Sr & CP2?  &	    &	       & C    & 	    &               \\
 GSC\,01870-01556    & 10302	&  86.39660 &	27.57710 & 10.50 &	   A0,Sr & CP2?  &	    &	       & B    & 	    &               \\
 GSC\,01870-01678    & 10308	&  86.45320 &	27.70060 & 11.10 &	   A0,Sr & CP2?  &	    &	       & B    & 	    &               \\
 GSC\,02405-00885    & 10317	&  86.60060 &	31.27390 & 10.70 &	A2,Si,Cr & CP2?  &	    &	       & B    & 	    &               \\
 AAO+24\,194	     & 10333	&  86.71400 &	24.51800 & 12.13 &	A2,Si,Sr & CP2?  &	    &	       & B    & 	    &               \\
 AAO+27\,185	     & 10365	&  86.93900 &	27.16750 & 11.95 &	   B9,Sr & CP2?  &   1.5409 &	0.0006 & BW   & 	    &               \\
 HD\,247664	     & 10385	&  86.98800 &	23.83660 & 10.40 &	B9,Si,Sr & CP2?  &	    &	       & B    & 	    &               \\
 HD\,247607	     & 10372	&  87.00810 &	29.82130 & 10.60 &	   B9,Sr & CP2?  &	    &	       & C    & 	    &               \\
 AAO+31\,260	     & 10373	&  87.03770 &	31.41910 & 11.16 &	B9,Si,Sr & CP2?  &	    &	       & C    & 	    &               \\
 HD\,248072	     & 10447	&  87.51170 &	23.67330 & 10.50 &	B9,Si,Cr & CP2?  &	    &	       & C    & 	    &               \\
 AAO+31\,290	     & 10449	&  87.62550 &	31.58220 & 11.60 &	   B9,Sr & CP2?  &	    &	       & B    & 	    &               \\
 GSC\,01875-00443    & 10572	&  88.31570 &	29.68850 & 11.00 &	   B9,Si & CP2?  &	    &	       & C    & 	    &               \\
 HD\,248815	     & 10596	&  88.42090 &	23.45250 & 10.57 &	   B9,Si & CP2?  &	    &	       & B    & 	    &               \\
 HD\,248767	     & 10591	&  88.42370 &	28.67700 & 10.13 &   A0,Si,Cr,Sr & CP2?  &	    &	       & C    & 	    &               \\
 HD\,248769	     & 10594	&  88.43610 &	27.41140 & 10.00 &	B8,Si,Sr & CP2?  &	    &	       & B    & 	    &               \\
 AAO+28\,312	     & 10595	&  88.46310 &	28.37830 & 10.93 &	B9,Si,Sr & CP2?  &	    &	       & B    & 	    &               \\
 HD\,248944	     & 10609	&  88.66830 &	28.07180 & 10.20 &	A1,Si,Sr & CP2?  &	    &	       & B    & 	    &               \\
 AAO+27\,276	     & 10633	&  88.90310 &	27.31990 & 10.94 &	   B8,Si & CP2?  &	    &	       & C    & 	    &               \\
 HD\,39865	     & 10643	&  89.12840 &	29.75200 &  8.61 &	   B8,Si & CP2?  &   0.9186 &	0.0002 & BW   & 	    &               \\
 AAO+30\,338	     & 10661	&  89.24500 &	30.90880 & 11.77 &	B8,Si,Sr & CP2?  &   3.4992 &	0.0027 & BW   & 	    &               \\
 GSC\,02406-01432    & 10662	&  89.26840 &	31.21150 & 10.90 &	   A0,Si & CP2?  &	    &	       & B    & 	    &               \\
 AAO+28\,354	     & 10664	&  89.28640 &	28.10530 &  9.56 &	   B9,Sr & CP2?  &	    &	       & B    & 	    &               \\
 GSC\,01867-02081    & 10729	&  89.69550 &	25.93460 & 11.50 &	B9,Si,Sr & CP2?  &	    &	       & C    & 	    &               \\
 HD\,250027	     & 10805	&  90.04370 &	28.57790 & 10.17 &	A1,Si,Sr & CP2?  &	    &	       & C    & S+W:19.7803 &               \\
 HD\,250149	     & 10825	&  90.20590 &	30.75830 & 10.50 &	B9,Si,Sr & CP2?  &	    &	       & C    & 	    &               \\
 HD\,41055	     & 10973	&  91.00220 &	28.64630 &  8.80 &	   B9,Si & CP2?  &	    &	       & C    & 	    &               \\
 GSC\,01868-00948    & 11023	&  91.21190 &	25.82570 & 11.50 &	B9,Si,Sr & CP2?  &	    &	       & C    & 	    &               \\
 HD\,41282	     & 11031	&  91.28880 &	24.34480 &  8.65 &	   B9,Si & CP2?  &	    &	       & B    & 	    &               \\
 HD\,251408	     & 11084	&  91.53450 &	27.81520 & 10.70 &   A1,Si,Cr,Sr & CP2?  &	    &	       & B    & 	    &               \\
 HD\,41418	     & 11087	&  91.56510 &	27.93470 &  8.64 &	   A0,Si & CP2?  &	    &	       & B    & 	    &               \\
 HD\,251621	     & 11105	&  91.66430 &	24.28540 &  9.97 &	   A0,Sr & CP2?  &	    &	       & B    & 	    &               \\
 GSC\,01872-02444    & 11101	&  91.67110 &	27.75790 & 10.50 &	A0,Si,Sr & CP2?  &	    &	       & B    & 	    &               \\
 HD\,251582	     & 11104	&  91.68080 &	26.17420 & 10.38 &	   A3,NA & CP2?  &	    &	       & B    & 	    &               \\
 HD\,251784	     & 11113	&  91.86190 &	26.02850 & 10.13 &	   A2,NA & CP2?  &	    &	       & C    & 	    &               \\
 HD\,251947	     & 11131	&  91.94900 &	23.19990 & 10.95 &	   A2,NA & CP2?  &	    &	       & B    & 	    &               \\
 HD\,251879	     & 11122	&  91.95200 &	26.32510 & 10.80 &	   B9,Sr & CP2?  &	    &	       & B    & 	    &               \\
 AAO+27\,454	     & 11136	&  92.03300 &	27.55900 & 11.67 &	A0,Si,Sr & CP2?  &	    &	       & B    & 	    &               \\
 HD\,252106	     & 11162	&  92.09440 &	23.62650 & 10.66 &	   A0,Si & CP2?  &	    &	       & B    & 	    &               \\
 GSC\,00742-02169    & 11180	&  92.10050 &	13.94660 & 10.80 &	   A0,Si & CP2   &   1.4828 &	0.0004 & BW   & C01:1.563   &               \\
 HD\,252104	     & 11168	&  92.12350 &	25.09050 & 10.50 &	B9,Si,Sr & CP2?  &	    &	       & B    & 	    &               \\
 AAO+27\,460	     & 11166	&  92.15390 &	27.62310 & 11.17 &	   A0,Si & CP2?  &	    &	       & C    & 	    &               \\
 HD\,41869	     & 11177	&  92.16520 &	25.65730 &  9.04 &	   B9,Si & CP2   &	    &	       & B    & 	    &               \\
 HD\,252459	     & 11230	&  92.44260 &	24.22580 & 10.50 &	    A,Si & CP2?  &	    &	       & B    & 	    &               \\
 AAO+24\,463	     & 11271	&  92.55130 &	24.14160 & 11.41 &	   B9,Si & CP2?  &	    &	       & C    & 	    &               \\
 AAO+24\,467	     & 11272	&  92.55850 &	24.41250 & 10.55 &	   B9,Si & CP2?  &	    &	       & B    & 	    &               \\
 BD+23\,1328	     & 11810	&  95.74010 &	23.27380 & 10.30 &	   A2,Sr & CP2   &	    &	       & C    & 	    &               \\
 HD\,44903	     & 11880	&  96.33670 &	23.05680 &  8.39 &	A5,Sr,Eu & CP2   &	    &	       & B    & 	    &               \\
 HD\,47103	     & 12630	&  99.43360 &	19.94860 &  9.15 &	 A,Sr,Eu & CP2   &	    &	       & C    & 	    &               \\
 HD\,263361	     & 13070	& 101.49600 &	19.16200 &  9.27 &	   B9,Si & CP2   &	    &	       & B    & 	    &               \\
 HD\,48953	     & 13123	& 101.70600 &	16.77250 &  6.81 &	 F,Sr,Eu & CP2   &   2.8939 &	0.0024 & BSW  & 	    &               \\
 HD\,50403	     & 13790	& 103.51500 &	22.26250 &  9.23 &	A2,Sr,Eu & CP2   &	    &	       & C    & 	    &               \\
 BD+23\,1580	     & 14530	& 105.97100 &	23.07520 &  9.83 &	A2,Cr,Eu & CP2?  &   2.4674 &	0.0019 & BW   & 	    &               \\
 GSC\,01909-01687    & 15650	& 110.32900 &	22.57300 & 10.50 &	   NA,Sr & CP2   &	    &	       & B    & 	    &               \\
 HD\,62510	     & 17130	& 116.28900 &	20.31600 &  6.54 &	   A0,Si & CP2?  &	    &	       & B    & 	    & 100.0(9.5/2)  \\
 HD\,72359	     & 20027	& 128.16600 &	10.06600 &  6.48 &	   A1,Sr & CP2?  &	    &	       & CS   & 	    & 19.0(4.0/4)   \\
 GSC\,01398-00532    & 20050	& 128.33501 &	20.40690 & 11.40 &	   NA,He & CP4?  &   3.9607 &	0.0053 & BW   & 	    &               \\
 HD\,94603	     & 27300	& 163.82100 &	-1.42467 &  9.35 &   A0,Sr,Eu,Cr & CP2   &	    &	       & C    & 	    &               \\
 HD\,96003	     & 27650	& 166.13901 &	12.66700 &  6.87 &	A3,Sr,Cr & CP2   &	    &	       & B    & 	    &               \\
 HD\,109860	     & 31890	& 189.51801 &	 3.28245 &  6.33 &	   A1,Si & CP2?  &	    &	       & CS   & 	    & 62.1(2.0/5)   \\
 HD\,112118	     & 32560	& 193.56799 &  -10.66760 & 10.23 &	A0,Cr,Eu & CP2?  &	    &	       & B    & 	    &               \\
 HD\,118054	     & 34040	& 203.66901 &  -13.21430 &  5.92 &   A1,Sr,Eu,Cr & CP2?  &   1.7912 &	0.0010 & BW   & 	    & 54.0(5.7/2)   \\
 HD\,125248	     & 35760	& 214.65900 &  -18.71600 &  5.85 &	A1,Eu,Cr & CP2   &   9.3105 &	0.0130 & SW   & C01:9.295   & 12.5(5.7/2)   \\
 HD\,126365	     & 36018	& 216.34900 &  -14.08500 &  8.47 &	F0,Sr,Cr & CP2   &	    &	       & C    & 	    &               \\
 HD\,138426	     & 39420	& 233.14200 &  -19.40270 &  8.57 &	B9,Sr,Cr & CP2   &   1.6571 &	0.0008 & SW   & 	    &               \\
 HD\,138633	     & 39460	& 233.39200 &  -11.06520 &  8.63 &   F0,Sr,Eu,Cr & CP2   &	    &	       & C    & 	    &               \\
 HD\,139160	     & 39630	& 234.36900 &  -26.27990 &  6.19 &    B7,He-weak & CP4?  &   6.2609 &	0.0218 & BW   & 	    & 50.0(1.3/5)   \\
 HD\,143517	     & 40675	& 240.39400 &  -21.72100 &  9.57 &	   A3,Sr & CP2   &	    &	       & B    & 	    &               \\
 HD\,144748	     & 41020	& 242.08701 &  -25.12720 &  8.60 &   F0,Sr,Eu,Cr & CP2   &   3.9815 &	0.0062 & BW   & 	    &               \\
 HD\,146254	     & 41360	& 243.96500 &  -14.84910 &  6.09 &	   A0,Si & CP2?  &	    &	       & C    & 	    & 137.5(14.3/2) \\
 HD\,146998	     & 41520	& 245.04100 &  -25.85730 &  9.65 &	A6,Sr,Cr & CP2   &	    &	       & C    & C98:1.2	    & 25.0(3.8/1)   \\
 HD\,148321	     & 41900	& 247.06300 &  -25.45400 &  6.97 &	A1-A8,Sr & CP2   &	    &	       & C    & 	    & 55.0(8.3/1)   \\
 HD\,148593	     & 42010	& 247.41400 &  -14.58510 &  9.13 &	   A2,Sr & CP2   &	    &	       & B    & 	    &               \\
 GSC\,06807-00646    & 42000	& 247.53200 &  -28.56320 & 10.70 &	   NA,Si & CP2   &	    &	       & C    & 	    &               \\
 HD\,149228	     & 42200	& 248.60100 &  -25.54970 & 10.06 &	   B9,Si & CP2   &	    &	       & C    & 	    &               \\
 HD\,150035	     & 42440	& 249.88200 &  -27.28560 &  8.71 &   A3,Cr,Eu,Sr & CP2   &   2.3389 &	0.0013 & W    & C98:0.54    & 75.0(11.3/1)  \\
 HD\,150715	     & 42645	& 250.94800 &  -25.21780 & 10.13 &   A0,Sr,Cr,Eu & CP2   &	    &	       & C    & 	    &               \\
 HD\,151941	     & 42980	& 252.90500 &  -26.68970 &  9.91 &   A7,Sr,Eu,Cr & CP2   &   0.0659 &	0.0001 & BW   & 	    &               \\
 HD\,191430	     & 53330	& 302.56201 &  -13.01010 &  8.41 &   A7,Sr,Eu,Cr & CP2   &	    &	       & B    & 	    &               \\
 HD\,191695	     & 53430	& 303.01001 &  -21.30080 &  9.87 &   A3,Sr,Eu,Cr & CP2   &	    &	       & B    & 	    &               \\
 HD\,196470	     & 54770	& 309.54099 &  -17.50180 &  9.72 &	A2,Sr,Eu & CP2   &	    &	       & B    & 	    &               \\
 HD\,206088	     & 57390	& 325.02301 &  -16.66230 &  3.68 &	A7-F3,Sr & CP2   &	    &	       & S    & 	    & 34.5(1.9/4)   \\
 HD\,206103	     & 57400	& 325.02499 &  -11.32830 &  9.47 &	   A0,Si & CP2   &	    &	       & C    & 	    &               \\
 HD\,207969	     & 57820	& 328.34000 &  -14.18930 &  8.16 &   A7,Sr,Eu,Cr & CP2   &	    &	       & CS   & 	    &               \\
 HD\,215766	     & 59600	& 341.92801 &  -14.05640 &  5.68 &	   B9,Si & CP2?  &   5.2310 &	0.0087 & BSW  & 	    & 80.0(12.0/1)  \\
 HD\,215913	     & 59640	& 342.16501 &	-2.15105 &  9.71 &   A0,Sr,Eu,Cr & CP2   &   1.8396 &	0.0010 & BSW  & 	    &               \\
\hline                                                  
\end{longtable}

%% file: tables/variablestableforpaper.tex
\begin{landscape}
\begin{table}
\tiny
\begin{center}
\caption{\label{tab:variable} Properties of the CP2 and CP4 stars identified as photometrically variable.}
\begin{tabular}{lrrrcrccccrrcccccc}
\hline
\hline       
Name & \# R09 & RA  & DEC & V  & Spectral & CP    & Period & $\sigma_{period}$ & MJD     & Analysis & Literature & $<Be>$ & \vsini & \Teff & \logl & \M    & $\tau$ \\
     &        & deg.& deg.& mag& Type     & class & days   & days              & maximum & remarks  & period     & G      & \kms   & K     & dex   &       &        \\
\hline
\hline
 HD\,315             & 30       &   1.93380 &   -2.54868 &  6.45 &         B9,Si & CP2   &   0.7584 &   0.0002 & 54155.923 &    * &                                       & R:1520 & 70(10/1)      & 12524(76/4)   & 2.12(0.14)   & 3.28(0.18) & 0.29(0.28) \\
 HD\,3992            & 1103     &  10.63670 &    5.67536 &  7.70 &         A3,Si & CP2   &   0.9898 &   0.0003 & 54167.709 &    B &                                       &        &               & 8924(49/50)   & 2.17(0.30/T) & 2.96(0.30) & 1(0)       \\
 HD\,5844            & 1543     &  15.03270 &    4.89430 &  9.71 &      A0,Si,Cr & CP2   &   1.8642 &   0.0012 & 54172.054 &      &                                       &        &               &               &              &            &            \\
 HD\,6164            & 1580     &  15.70570 &    7.50076 &  7.85 &         A0,Si & CP2   &   3.7931 &   0.0048 & 54216.332 &      &                                       & K:90   &               & 10107(323/3)  & 2.19(0.39)   & 3.10(0.48) & 0.88(0.21) \\
 HD\,6397            & 1637     &  16.27230 &   14.94610 &  5.65 &         F3,Sr & CP2   &   4.4162 &   0.0063 & 54190.326 &    B &                                       &        & 5.8(0.6/2)    & 6775(399/3)   & 1.20(0.09)   & 1.81(0.08) & 0.89(0.13) \\
 HD\,10783           & 2670     &  26.42720 &    8.55923 &  6.58 &   A2,Si,Cr,Sr & CP2   &   4.1321 &   0.0037 & 54188.811 &   B* &C01:4.133 / W:4.1327 / B:4.14628       & B:1269 & 25.0(3.8/1)   & 10345(208/4)  & 1.86(0.13)   & 2.70(0.14) & 0.58(0.21) \\
 $\gamma$\,Ari       & 2930     &  28.38240 &   19.29410 &  3.89 &   A1,Si,Cr,Sr & CP2   &   1.6092 &   0.0005 & 54187.512 &  BS* &C98:1.6092 / B:1.6093                  & B:545  & 115.0(5.7/2)  & 9958(281/5)   & 1.80(0.11)   & 2.61(0.12) & 0.62(0.18) \\
 HD\,12447           & 3180     &  30.51180 &    2.76376 &  3.83 &   A2,Si,Sr,Cr & CP2   &   0.7455 &   0.0001 & 54187.669 &   BS &C01:1.4909 / W:0.745483 / B:1.4907     & B:365  & 49.0(2.7/2)   & 10229(253/3)  & 1.83(0.11)   & 2.66(0.13) & 0.58(0.20) \\
                     &          &           &            &       &               &       &   6.6507 &   0.0195 &           &      &                                       &        &               &               &              &            &            \\
 HD\,19832           & 4910     &  48.05940 &   27.25700 &  5.77 &         B8,Si & CP2   &   0.7279 &   0.0001 & 54206.482 &    * &C01:0.7279 / S+W:0.727902 / B:0.727893 & B:315  & 103.3(5.8/3)  & 12430(360/99) & 2.24(0.17)   & 3.41(0.22) & 0.51(0.26) \\
 HD\,20629           & 5160     &  49.94900 &   19.07620 &  7.44 &   A0,Si,Cr,Sr & CP2   &   2.4994 &   0.0018 & 54207.068 &   B* &C01:2.5 / S+W:2.4997                   &        &               & 12500(27/3)   & 2.04(0.30)   & 3.19(0.33) & 0.17(0.31) \\
 HD\,21590           & 5400     &  52.42740 &   16.76230 &  7.07 &         B9,Si & CP2   &   2.7609 &   0.0014 & 54210.073 &    * &                                       & B:1098 &               & 12199(322/4)  & 1.82(0.14/Z) & 2.92(0.15) & 0(0)       \\
 HD\,22860           & 5830     &  55.32660 &   28.70280 &  6.89 &         B9,Si & CP2   &   6.2685 &   0.0120 & 54212.844 &      &                                       &        &               & 10876(-/1)    & 1.83(0.15)   & 2.74(0.16) & 0.38(0.30) \\
 HD\,24155           & 6190     &  57.81610 &   13.04610 &  6.31 &         B9,Si & CP2   &   2.5345 &   0.0020 & 54213.356 &    * &C01:2.535 / S+W:2.53465 / B:2.53465    & B:1034 & 47.5(3.7/2)   & 13780(70/99)  & 2.04(0.11/Z) & 3.40(0.15) & 0(0)       \\
 HD\,24769           & 6330     &  59.26590 &   23.17550 &  6.06 &         B9,Si & CP2?  &   1.4876 &   0.0005 & 54215.816 &   B* &C98:1.49 / C01:2.975                   &        & 70.0(6.4/2)   & 9617(159/5)   & 2.24(0.16)   & 3.14(0.21) & 0.96(0.07) \\
 HD\,27309           & 7010     &  64.90290 &   21.77350 &  5.38 &      A0,Si,Cr & CP2   &   1.5689 &   0.0007 & 54220.907 &   B* &C01:1.569 / S+W:1.56896 / B:1.10496    & B:1755 & 44.3(4.6/3)   & 11930(250/99) & 1.96(0.10)   & 3.01(0.13) & 0.22(0.23) \\
 HD\,27404           & 7030     &  65.15740 &   28.89200 &  7.97 &         A0,Si & CP2   &   2.7793 &   0.0009 & 54221.481 &    * &C01:2.7793 / W:2.77929                 & R:1700 &               & 10954(160/3)  & 1.96(0.31)   & 2.88(0.35) & 0.53(0.41) \\
 HD\,30466           & 7870     &  72.31670 &   29.57140 &  7.28 &         A0,Si & CP2   &   1.4069 &   0.0006 & 54227.030 &   B* &C01:1.4 / S+W:1.39 / B:1.39            & B:1465 &               & 11081(280/4)  & 1.64(0.15/Z) & 2.59(0.14) & 0(0)       \\
 HD\,32549           & 8280     &  76.14230 &   15.40410 &  4.68 &      B9,Si,Cr & CP2   &   4.6397 &   0.0065 & 54229.586 &   B* &C01:4.64 / S+W:4.6398                  & B:50   & 34.3(2.6/3)   & 9730(-/99)    & 2.17(0.10)   & 3.05(0.13) & 0.91(0.08) \\
 HD\,32576           & 8290     &  76.17420 &   14.85900 &  6.78 &         A3,NA & CP2?  &   6.3844 &   0.0075 & 54234.341 &   BS &                                       &        &               & 8657(20/50)   & 1.16(0.11)   & 1.92(0.10) & 0.15(0.24) \\
 HD\,32643           & 8340     &  76.31900 &   15.23990 &  7.69 &         A0,NA & CP2?  &   2.3442 &   0.0017 & 54229.550 &    B &                                       &        &               & 8615(230/50)  & 2.00(0.31)   & 2.76(0.32) & 0.97(0.10) \\
 HD\,34547           & 8820     &  79.72050 &   13.56720 &  7.46 &      A0,Sr,Cr & CP2   &   1.7145 &   0.0010 & 54232.115 &    * &                                       &        &               & 10077(-/1)    & 1.22(0.17/Z) & 2.16(0.13) & 0(0)       \\
 HD\,34719           & 8850     &  80.07620 &   19.57820 &  6.67 &   A0,Hg,Si,Cr & CP2   &   1.6401 &   0.0007 & 54233.114 &    * &C01:1.6399 / W:1.63988                 & K:880  & 57.0(11.0/1)  & 11933(231/4)  & 2.05(0.14)   & 3.11(0.17) & 0.38(0.28) \\
 HD\,242764          & 8872     &  80.32910 &   32.13500 &  9.89 &      B8,Si,Sr & CP2?  &   5.1253 &   0.0062 & 54234.014 &    B &                                       &        &               &               &              &            &            \\
 HD\,35033           & 8948     &  80.77380 &   31.15850 &  9.05 &         A0,Si & CP2?  &   3.9575 &   0.0048 & 54235.800 &    B &                                       &        &               & 12880(-/1)    &              & 4.01(0.80) &            \\
 HD\,35379           & 9053     &  81.40870 &   30.68730 &  8.35 &      B9,Si,Sr & CP2?  &   2.5154 &   0.0019 & 54235.297 &    B &                                       &        &               & 14096(-/1)    &              & 4.60(0.92) &            \\
 HD\,35533           & 9145     &  81.51120 &   15.67180 &  7.56 &         B9,Si & CP2   &   2.0641 &   0.0011 & 54234.113 &   B* &                                       &        &               & 12565(-/1)    & 1.78(0.16/Z) & 2.96(0.16) & 0(0)       \\
 HD\,35479           & 9090     &  81.56090 &   29.98670 &  8.09 &         B9,Si & CP2   &   0.9660 &   0.0003 & 54235.136 &    B &                                       &        &               & 11997(161/2)  & 1.75(0.21/Z) & 2.83(0.19) & 0(0)       \\
 HD\,245687          & 9993     &  84.65980 &   27.36490 & 10.36 &         B8,Si & CP2?  &   1.4700 &   0.0007 & 54238.548 &    B &                                       &        &               &               &              &            &            \\
 HD\,246706          & 10232    &  85.90340 &   29.75280 &  9.12 &      B8,Si,Sr & CP2?  &   1.4697 &   0.0006 & 54238.529 &    * &                                       &        & 35.0(5.3/1)   & 12561(3/2)    & 1.59(0.37/Z) & 2.81(0.27) & 0(0)       \\
 HD\,246726          & 10236    &  85.97110 &   30.73380 & 10.61 &      A5,Si,Sr & CP2?  &   5.8212 &   0.0085 & 54242.133 &    B &                                       &        &               &               &              &            &            \\
 HD\,39135           & 10507    &  88.00370 &   33.55360 &  9.07 &   A0,Si,Cr,Sr & CP2?  &   3.9091 &   0.0038 & 54284.992 &    B &                                       &        &               &               &              &            &            \\
 HD\,39317           & 10560    &  88.09290 &   14.17180 &  5.58 &   B9,Si,Eu,Cr & CP2   &   2.6569 &   0.0022 & 54241.530 &   B* &C98:2.65 / S+W:2.6541                  & B:400  & 35.0(2.6/3)   & 9704(92/5)    & 1.94(0.10)   & 2.74(0.12) & 0.79(0.12) \\
 HD\,250179          & 10847    &  90.15640 &   22.96030 & 11.20 &         B9,Si & CP2?  &   0.9505 &   0.0005 & 54241.816 &  BS* &                                       &        &               &               &              &            &            \\
 HD\,40696           & 10877    &  90.38120 &   23.73870 &  8.24 &         A0,Si & CP2?  &   0.9828 &   0.0005 & 54242.390 &   B* &                                       &        &               & 9122(116/2)   &              & 2.51(0.49) &            \\
 HD\,40833           & 10917    &  90.60800 &   24.77460 &  9.28 &         B9,Si & CP2   &   3.4227 &   0.0033 & 54243.083 &    B &                                       &        &               & 11342(-/1)    &              & 3.32(0.64) &            \\
 GSC\,01864-00336    & 10945    &  90.75950 &   22.82520 & 10.85 &      B8,Si,Sr & CP2?  &   0.9479 &   0.0003 & 54242.569 &   BS &                                       &        &               &               &              &            &            \\
 HD\,43819           & 11620    &  94.75770 &   17.32530 &  6.29 &         B9,Si & CP2   &  15.0285 &   0.0476 & 54254.609 &  BS* &C01:15.03                              & B:269  & 23.3(5.8/3)   & 10930(290/99) & 2.36(0.25)   & 3.42(0.38) & 0.87(0.16) \\
                     &          &           &            &       &               &       &   5.5697 &   0.0078 &           &      &                                       &        &               &               &              &            &            \\
 HD\,44636           & 11830    &  95.85550 &   15.84550 &  8.99 &   B8,Cr,Eu,Si & CP2   &   2.6311 &   0.0017 & 54247.088 &    B &                                       & K:360  &               & 11371(-/1)    &              & 3.33(0.65) &            \\
 HD\,44738           & 11850    &  95.97510 &   14.11350 &  7.92 &         B9,Si & CP2   &   1.5764 &   0.0004 & 54248.572 &   B* &                                       &        &               & 10429(394/3)  & 2.00(0.40)   & 2.88(0.50) & 0.70(0.40) \\
 HD\,46593           & 12470    &  98.73640 &   16.76380 &  7.22 &         B9,NA & CP2   &   2.8691 &   0.0018 & 54250.734 &    * &                                       &        &               & 13297(-/1)    & 2.34(0.21)   & 3.68(0.28) & 0.43(0.34) \\
 HD\,47152           & 12670    &  99.59590 &   28.98440 &  5.77 &   A0,Eu,Cr,Hg & CP2   &   2.7112 &   0.0020 & 54251.340 &    * &C98:2.8                                &        & 58.3(4.6/3)   & 9715(188/6)   & 1.72(0.12)   & 2.49(0.13) & 0.59(0.21) \\
 HD\,50341           & 13780    & 103.55700 &   33.00250 &  8.20 &   B9,Sr,Cr,Eu & CP2   &   2.5096 &   0.0023 & 54688.505 &    * &C01:2.5092 / W:2.50919                 &        &               & 9862(-/1)     & 1.48(0.24)   & 2.29(0.20) & 0.18(0.30) \\
 HD\,52181           & 14330    & 105.33800 &   20.09780 &  8.96 &         A8,Sr & CP2   &   2.9176 &   0.0034 & 54257.226 &    B &                                       &        &               & 8516(90/50)   &              & 2.33(0.47) &            \\
 HD\,55579           & 15120    & 108.61100 &   24.71110 &  6.89 &         A0,Sr & CP2   &   0.6970 &   0.0001 & 54257.937 &   B* &C01:2.315 / KE:0.697054                &        &               & 9751(187/6)   & 1.65(0.14)   & 2.43(0.14) & 0.49(0.26) \\
 HD\,68351           & 18900    & 123.28700 &   29.65650 &  5.62 &      A0,Si,Cr & CP2   &   3.3095 &   0.0029 & 54272.391 &   S* &C98:3.2                                & B:152  & 27.5(5.7/2)   & 9925(59/4)    & 2.35(0.10)   & 3.33(0.14) & 0.98(0.08) \\
 HD\,74521           & 20790    & 131.18800 &   10.08170 &  5.66 &   A1,Si,Eu,Cr & CP2   &   6.9071 &   0.0116 & 54286.485 &   S* &C01:7.05 / S+W:4.2359 / B:7.76851      & B:812  & 19.0(4.6/3)   & 10789(352/4)  & 1.99(0.11)   & 2.89(0.13) & 0.61(0.18) \\
 HD\,77350           & 21860    & 135.68401 &   24.45290 &  5.47 &   B9,Sr,Cr,Hg & CP2   &   4.2025 &   0.0040 & 54281.001 &    S &C98:4.19 / B:4.024                     & B:846  & 18.9(0.4/5)   & 10250(-/99)   & 1.97(0.11)   & 2.82(0.13) & 0.70(0.15) \\
                     &          &           &            &       &               &       &   2.8024 &   0.0015 &           &      &                                       &        &               &               &              &            &            \\
                     &          &           &            &       &               &       &   2.6511 &   0.0013 &           &      &                                       &        &               &               &              &            &            \\
 HD\,90569           & 26010    & 156.91200 &    9.76240 &  6.03 &   A0,Sr,Cr,Si & CP2   &   3.0525 &   0.0031 & 54304.722 &    * &S+W:7.897 / B:1.445                    & B:192  & 30.0(4.5/1)   & 10500(-/99)   & 1.81(0.10)   & 2.67(0.12) & 0.47(0.22) \\
 HD\,97859           & 28210    & 168.87900 &    4.95653 &  9.36 &         B9,Si & CP2   &   0.7921 &   0.0002 & 54315.528 &      &                                       & B:400  &               & 13711(232/4)  &              & 4.41(0.88) &            \\
 HD\,99665           & 28683    & 172.01500 &   -0.89786 &  7.13 &      A0,Sr,Eu & CP2?  &   1.9835 &   0.0008 & 54321.036 &    S &                                       &        &               & 9753(35/2)    & 1.32(0.2/Z)  & 2.17(0.15) & 0(0)       \\
                     &          &           &            &       &               &       &   4.2358 &   0.0021 &           &      &                                       &        &               &               &              &            &            \\
 HD\,107000          & 30960    & 184.58000 &    3.10112 &  8.01 &         A2,Sr & CP2   &   2.8187 &   0.0024 & 54332.587 &    * &                                       & R:200  &               & 8922(262/50)  & 1.96(0.36)   & 2.73(0.39) & 0.91(0.17) \\
 HD\,107452          & 31140    & 185.28500 &  -11.47580 &  8.62 &   A7,Sr,Cr,Eu & CP2   &   2.1462 &   0.0011 & 54336.618 &    B &                                       &        &               & 8001(188/2)   & 0.32(0.21/Z) & 1.67(0.21) & 0(0)       \\
 HD\,111133          & 32310    & 191.75999 &    5.95037 &  6.34 &   A1,Sr,Cr,Eu & CP2   &   2.2246 &   0.0017 & 55025.192 &    S &C01:16.31 / S+W:16.304 / B:16.3078     & B:806  & 10.0(1.1/2)   & 9850(220/99)  & 2.27(0.16)   & 3.20(0.22) & 0.95(0.08) \\
                     &          &           &            &       &               &       &  15.9490 &   0.1876 &           &      &                                       &        &               &               &              &            &            \\
 HD\,116114          & 33530    & 200.44299 &  -18.74210 &  7.04 &   F0,Sr,Cr,Eu & CP2?  &   5.3832 &   0.0100 & 54352.057 &    B &                                       & B:1923 & 65.0(9.8/1)   & 7850(210/99)  & 1.39(0.13)   & 2.04(0.12) & 0.75(0.16) \\
 HD\,130158          & 37080    & 221.84399 &  -25.62430 &  5.61 &         B9,Si & CP2   &   4.4136 &   0.0032 & 54374.695 &    * &                                       & B:350  & 55.0(4.3/3)   & 10462(196/4)  & 2.50(0.13)   & 3.61(0.16) & 1.00(0.05) \\
 HD\,130559          & 37160    & 222.32899 &  -14.14900 &  5.32 &   A1,Sr,Cr,Eu & CP2   &  25.3992 &   0.1970 & 54385.869 &   S* &                                       & B:1375 & 29.0(1.7/3)   & 9592(260/6)   & 1.62(0.11)   & 2.39(0.12) & 0.51(0.23) \\
                     &          &           &            &       &               &       &   1.8871 &   0.0008 &           &      &                                       &        &               &               &              &            &            \\
 HD\,134759          & 38230    & 228.05499 &  -19.79170 &  4.54 &         B9,Si & CP2   &   3.0993 &   0.0012 & 54376.576 &    * &                                       & B:320  & 51.0(5.8/3)   & 11700(108/4)  & 2.44(0.10)   & 3.61(0.15) & 0.83(0.09) \\
                     &          &           &            &       &               &       &   0.3000 &   0.0050 &           &      &                                       &        &               &               &              &            &            \\
                     &          &           &            &       &               &       &   5.8395 &   0.0097 &           &      &                                       &        &               &               &              &            &            \\
 HD\,137916          & 39210    & 232.35201 &  -19.40470 &  8.74 &         B9,Si & CP2   &   3.8827 &   0.0023 & 54380.984 &   B* &                                       &        &               & 10493(-/1)    &              & 2.99(0.57) &            \\
 HD\,137949          & 39240    & 232.39500 &  -17.44090 &  6.68 &   F0,Sr,Eu,Cr & CP2   &   4.8511 &   0.0080 & 54379.311 &   BS &C01:6.68 / B:11.13313                  & B:1497 &               & 7530(40/99)   & 1.13(0.12)   & 1.79(0.10) & 0.63(0.23) \\
\hline
\multicolumn{18}{l}{\emph{continued on next page}}
\end{tabular}
\end{center}
\end{table}
\end{landscape}
\begin{landscape}
\begin{table}
\setcounter{table}{2}
\tiny
\begin{center}
\caption{{\it Continued}}
\begin{tabular}{lrrrcrccccrrcccccc}
\hline
\hline
 HD\,141249          & 40104    & 237.16800 &  -18.41120 & 10.19 &      A3,Sr,Eu & CP2   &   1.2297 &   0.0004 & 54383.641 &    B &                                       &        &               & 8210(34/50)   &              & 2.23(0.46) &            \\
 HD\,142096          & 40340    & 238.33400 &  -20.16700 &  5.03 &    B3,He-weak & CP4   &   3.3136 &   0.0032 & 54387.383 &    S &                                       &        & 175.0(13.4/3) & 16478(131/4)  & 2.62(0.14/Z) & 4.65(0.27) & 0(0)       \\
                     &          &           &            &       &               &       &   0.5836 &   0.0001 &           &      &                                       &        &               &               &              &            &            \\
 HD\,142301          & 40380    & 238.66499 &  -25.24370 &  5.87 & B8,He-weak,Si & CP2   &   1.4596 &   0.0004 & 54387.047 &   B* &C01:1.4595 / S+W:1.45937 / B:1.45955   & B:2103 & 68.7(4.3/3)   & 15860(150/99) & 2.62(0.13)   & 4.54(0.22) & 0.16(0.22) \\
 HD\,142884          & 40510    & 239.45300 &  -23.52730 &  6.78 &         B9,Si & CP2   &   0.8030 &   0.0002 & 54389.202 &   S* &                                       & B:285  & 155.0(18.4/2) & 14223(270/4)  & 2.17(0.14/Z) & 3.63(0.19) & 0(0)       \\
 HD\,142990          & 40530    & 239.64500 &  -24.83150 &  5.43 &    B6,He-weak & CP4   &   0.9789 &   0.0003 & 54387.177 &    * &C01:0.979 / S:0.492 / B:0.9791         & B:1304 & 153.3(10.8/3) & 17700(1130/99)& 2.97(0.10)   & 5.55(0.28) & 0.30(0.27) \\
 HD\,145102          & 41120    & 242.56599 &  -26.90910 &  6.59 &         B9,Si & CP2   &   1.4178 &   0.0006 & 54390.576 &    * &                                       & B:281  & 71.3(1.5/4)   & 10919(182/4)  & 2.03(0.15)   & 2.95(0.18) & 0.63(0.21) \\
 HD\,145501          & 41240    & 242.99400 &  -19.45020 &  6.26 &         B9,Si & CP2   &   0.5853 &   0.0001 & 54388.737 &    B &                                       & B:1242 & 60.0(9.0/1)   & 13715(156/2)  &              & 4.41(0.88) &            \\
                     &          &           &            &       &               &       &   0.7323 &   0.0001 &           &      &                                       &        &               &               &              &            &            \\
 HD\,145792          & 41280    & 243.43900 &  -24.42240 &  6.40 &    B6,He-weak & CP4?  &   1.6956 &   0.0005 & 54389.824 &   B* &W:0.8478                               & B:286  & 25.7(2.3/3)   & 14710(143/4)  & 2.33(0.15/Z) & 3.90(0.23) & 0(0)       \\
 HD\,146001          & 41310    & 243.72301 &  -25.47700 &  6.06 &    B8,He-weak & CP4?  &   3.9146 &   0.0041 & 54392.197 &   B* &                                       & B:647  & 147.5(9.5/2)  & 13790(300/99) & 2.41(0.12)   & 3.84(0.21) & 0.40(0.29) \\
                     &          &           &            &       &               &       &   0.5577 &   0.0001 &           &      &                                       &        &               &               &              &            &            \\
 HD\,147010          & 41530    & 245.02299 &  -20.05640 &  7.40 &   B9,Si,Cr,Sr & CP2   &   3.9209 &   0.0048 & 54394.083 &   B* &C01:3.9207 / B:3.920676                & B:4032 & 26.5(2.7/2)   & 12752(219/4)  & 2.00(0.18/Z) & 3.18(0.20) & 0(0)       \\
 HD\,147105          & 41560    & 245.20500 &  -25.39430 &  8.82 &   A3,Sr,Cr,Eu & CP2   &   2.0025 &   0.0008 & 54392.388 &   B* &C98:1.9                                & B:456  & 50.0(7.5/1)   & 8200(269/3)   &              & 2.23(0.45) &            \\
 HD\,147890          & 41720    & 246.41299 &  -29.40040 &  7.67 &      A0,Si,Sr & CP2   &   4.3359 &   0.0059 & 54395.777 &    * &C01:4.336                              & B:235  & 46.7(2.6/3)   & 11329(300/4)  & 2.63(0.38)   & 3.90(0.53) & 0.97(0.13) \\
 HD\,150714          & 42640    & 250.91499 &  -22.73650 &  7.59 &         A0,Si & CP2   &   1.6290 &   0.0008 & 54396.038 &   B* &C01:1.6288                             &        &               & 10547(60/4)   & 1.71(0.19)   & 2.58(0.18) & 0.30(0.31) \\
 HD\,150716          & 42650    & 250.96400 &  -25.77780 &  9.70 &         A0,Si & CP2   &   1.9899 &   0.0013 & 54396.727 &   B* &                                       &        &               & 11368(42/2)   &              & 3.33(0.64) &            \\
 HD\,151346          & 42810    & 251.94400 &  -23.97430 &  7.91 &    B7,He-weak & CP4   &   2.1797 &   0.0017 & 54398.734 &      &                                       & B:245  & 46.0(9.0/1)   & 14404(187/2)  & 2.40(0.21)   & 3.94(0.31) & 0.19(0.29) \\
 HD\,190576          & 52978    & 301.55801 &  -19.49040 &  7.64 &         B9,Si & CP2   &   0.5063 &   0.0001 & 54095.403 &    B &                                       &        &               & 12352(-/1)    &              & 3.76(0.74) &            \\
 HD\,191287          & 53290    & 302.42499 &  -18.34670 &  8.22 &         B9,Eu & CP2   &   1.6235 &   0.0007 & 54096.503 &    * &C01:1.6234                             &        &               & 11445(-/1)    & 2.08(0.37)   & 3.08(0.45) & 0.57(0.43) \\
 HD\,199728          & 55630    & 314.90100 &  -19.03530 &  6.27 &         B9,Si & CP2   &   2.2411 &   0.0011 & 54112.643 &    * &C01:2.241 / S:2.25                     & R:400  & 57.5(5.8/2)   & 11935(114/3)  & 2.05(0.13)   & 3.11(0.16) & 0.37(0.27) \\
 HD\,207188          & 57640    & 326.90201 &  -17.29470 &  7.64 &         A0,Si & CP2   &   2.6733 &   0.0022 & 54124.810 &    * &S:2.67                                 & K:1220 & 30.0(4.5/1)   & 12303(35/3)   & 1.78(0.22/Z) & 2.91(0.20) & 0(0)       \\
 HD\,210424          & 58510    & 332.65601 &  -11.56490 &  5.44 &         B6,Si & CP2   &   3.9613 &   0.0033 & 54128.906 &    B &                                       &        & 20.0(2.1/2)   & 12662(257/3)  & 2.30(0.11)   & 3.53(0.16) & 0.54(0.19) \\
 HD\,211099          & 58630    & 333.72501 &   -6.73547 &  7.64 &         B9,Si & CP2   &   3.7887 &   0.0042 & 54132.431 &    * &                                       &        &               & 12396(106/3)  & 2.37(0.39)   & 3.59(0.58) & 0.67(0.42) \\
 HD\,220825          & 60520    & 351.73300 &    1.25561 &  4.95 &   A1,Cr,Sr,Eu & CP2   &   1.4150 &   0.0005 & 54149.117 &    * &C01:1.41 / S:0.58525 / B:1.14077       & B:269  & 35.0(4.6/3)   & 9200(80/99)   & 1.32(0.09)   & 2.09(0.09) & 0.20(0.23) \\
 HD\,224103          & 61430    & 358.78201 &    7.07097 &  6.22 &         A0,Si & CP2?  &   2.5341 &   0.0023 & 54503.094 &      &                                       &        & 25.0(5.7/2)   & 10460(156/4)  & 1.72(0.11)   & 2.58(0.13) & 0.36(0.28) \\
\hline                                                  
\end{tabular}
\end{center}
\end{table}
\smallskip

\noindent
B(period): \citet{bychkov2005}\\
B(magnetic field): \citet{bychkov2009}\\
C98: \citet{catalano1998}\\ 
C01: \citet{renson01}\\
K: \citet{kudryavtsev08}\\
KE: \citet{koen02}\\
K07: \citet{kraus}\\
R: \citet{romanyuk08}\\
S: \citet{samus09}\\
W: \citet{watson2006}
\end{landscape}

%% file: stereoCP-v5.0.bbl
\begin{thebibliography}{}
\bibitem[\protect\citeauthoryear{Adelman}{2008}]{adelman2008}
	Adelman, S.~J. 2008, PASP, 120, 595
\bibitem[\protect\citeauthoryear{Arenou et al.}{1992}]{arenou92}
	Arenou, F., Grenon, M. \& Gomez, A. 1992, A\&A, 258, 104
\bibitem[\protect\citeauthoryear{Babel}{1992}]{babel1992}
	Babel, J. 1992, A\&A, 258, 449
\bibitem[\protect\citeauthoryear{Bagnulo et al.}{2006}]{bagnulo06}
	Bagnulo, S., Landstreet, J.~D., Mason, E., Andretta, V., Silaj, J. 
	\& Wade, G.~A. 2006, A\&A, 450, 777
\bibitem[\protect\citeauthoryear{Bewsher et al.}{2010}]{bewsher2010}
	Bewsher, D., Brown, D.~S., Eyles, C.~J., Kellett, B.~J., White, G.~J.
	\& Swinyard, B. 2010, Solar Physics, 264, 433
\bibitem[\protect\citeauthoryear{Borra et al.}{1983}]{borra83}
	Borra, E.~F., Landstreet, J.~D. \&, Thompson, I. 1983, ApJS, 53, 151
\bibitem[\protect\citeauthoryear{Brown et al.}{2009}]{brown2009}
	Brown, D.~S., Bewsher, D. \& Eyles, C.~J. 2009, Solar Physics, 254, 185
\bibitem[\protect\citeauthoryear{Bychkov et al.}{2009}]{bychkov2009}
	Bychkov, V.~D., Bychkova, L.~V. \& Madej, J. 2009, MNRAS, 394, 1338
\bibitem[\protect\citeauthoryear{Bychkov et al.}{2005}]{bychkov2005}
	Bychkov, V.~D., Bychkova, L.~V. \& Madej, J. 2005, A\&A, 430, 1143
\bibitem[\protect\citeauthoryear{Catalano \& Renson}{1998}]{catalano1998}
	Catalano, F.~A. \& Renson, P. 1998, A\&AS, 127, 421
\bibitem[\protect\citeauthoryear{Cutri et al.}{2003}]{cutri03}
	Cutri, R.~M., Skrutskie, M.~F., van Dyk, S., et al., 2003, 2MASS All 
	Sky Catalog of point sources. The IRSA 2MASS All-Sky Point Source 
	Catalog, NASA/IPAC Infrared Science Archive. 
\bibitem[\protect\citeauthoryear{Eyles et al.}{2009}]{eyles2009}
	Eyles, C.~J., Harrison, R.~A., Davis, C.~J., et al. 2009, Solar 
	Physics, 254, 387
\bibitem[\protect\citeauthoryear{Glebocki \& Gnacinski}{2005}]{glebocki05} 
	Glebocki, R., \& Gnacinski, P. 2005, ESA, SP-560, 571, Vsini values
\bibitem[\protect\citeauthoryear{Gomez et al.}{1998}]{gomez98} 
	Gomez, A.~E., Luri, X., Grenier, S., Figueras, F., North, P., 
	Royer, F., Torra, J. \& Mennessier, M.~O. 1998, A\&A, 336, 953
\bibitem[\protect\citeauthoryear{Harmanec \& Bo{\v z}i{\'c}}{2001}]{harmanec01}
	Harmanec, P. \& Bo{\v z}i{\'c}, H. 2001, A\&A, 369, 1140
\bibitem[\protect\citeauthoryear{Horch et al.}{2004}]{horch04}
	Horch, E.~P., Meyer, R.~D. \& van Altena, W.~F. 2004, AJ, 127, 1727
\bibitem[\protect\citeauthoryear{Kaiser et al.}{2008}]{kaiser08}
	Kaiser, M.~L., Kucera, T.~A., Davila, J.~M., St.~Cyr, O.~C., 
	Guhathakurta, M. \& Christian, E. 2008, Space Science Reviews, 136, 5
\bibitem[\protect\citeauthoryear{Kharadze \& Chargeishvili}{1990}]{kharadze90}
	Kharadze, E.~K. \& Chargeishvili, K.~B. 1990, AJ, 379, 395
\bibitem[\protect\citeauthoryear{Kharchenko}{2001}]{kharchenko01}
	Kharchenko, N.~V. 2001, Kinematika i Fizika Nebesnykh Tel, 17, 409
\bibitem[\protect\citeauthoryear{Kochukhov \& Bagnulo}{2006}]{kb}
	Kochukhov, O. \& Bagnulo, S. 2006, A\&A, 450, 775
\bibitem[\protect\citeauthoryear{Koen \& Eyer}{2002}]{koen02}
	Koen, C. \& Eyer, L. 2002, MNRAS, 331, 45
\bibitem[\protect\citeauthoryear{Kraus et al.}{2007}]{kraus}
	Kraus, A.~L., Craine, E.~R., Giampapa, M.~S., Scharlach, W.~W.~G. \&
	Tucker, R.~A. 2007, AJ, 134, 1488
\bibitem[\protect\citeauthoryear{Kudryavtsev et al.}{2008}]{kudryavtsev08}
	Kudryavtsev, D.~O., Romanyuk, I.~I., Elkin, V.~G. \& Paunzen, E. 2006,
	MNRAS, 372, 1804
\bibitem[\protect\citeauthoryear{Kurtz}{1982}]{kurtz1982}
	Kurtz, D.~W. 1982, MNRAS, 200, 807
\bibitem[\protect\citeauthoryear{Kurtz}{1989}]{kurtz1989}
	Kurtz, D.~W. 1989, MNRAS, 238, 1077
\bibitem[\protect\citeauthoryear{Landstreet et al.}{2007}]{landstreet07}
	Landstreet, J.~D., Bagnulo, S., Andretta, V., Fossati, L., Mason, E., 
	Silaj, J. \& Wade, G.~A. 2007, A\&A, 470, 685
\bibitem[\protect\citeauthoryear{Mason et al.}{2007}]{mason07}
	Mason, B.~D., Hartkopf, W.~I., Wycoff, G.~L. \& Wieder, G. 2007, AJ,
	134, 1671
\bibitem[\protect\citeauthoryear{Mermilliod et al.}{1997}]{mermilliod97}
	Mermilliod, J.-C., Mermilliod, M. \& Hauck, B. 1997, A\&AS, 124, 349
\bibitem[\protect\citeauthoryear{Mikul{\'a}{\v s}ek}{2007}]{mikulaek2007}
	Mikul{\'a}{\v s}ek, Z. 2007, Astronomical and Astrophysical 
	Transactions, 26, 63
\bibitem[\protect\citeauthoryear{Netopil et al.}{2008}]{netopil08}
	Netopil, M., Paunzen, E., Maitzen, H.~M., North, P. \& Hubrig, S. 2008,
	A\&A, 491, 545
\bibitem[\protect\citeauthoryear{North}{1998}]{north98}
	North, P. 1998, A\&A, 334, 181
\bibitem[\protect\citeauthoryear{Paunzen et al.}{2011}]{paunzen2011} 
	Paunzen, E., Netopil, M., Pintado, O. I., Rode-Paunzen, M. 2011, 
	AN, 332, 77
\bibitem[\protect\citeauthoryear{Pourbaix et al.}{2004}]{sb9}
	Pourbaix, D., Tokovinin, A.~A., Batten, A.~H., et al.
	2004, A\&A, 424, 727
\bibitem[\protect\citeauthoryear{Preston}{1974}]{preston1974}
	Preston, G.\,W. 1974, ARA\&A, 12, 257
\bibitem[\protect\citeauthoryear{Renson \& Catalano}{2001}]{renson01}
	Renson, P. \& Catalano, F.~A. 2001, A\&A, 378, 113
\bibitem[\protect\citeauthoryear{Renson \& Manfroid}{2009}]{renson09}
	Renson, P. \& Manfroid, J. 2009, A\&A, 498, 961
\bibitem[\protect\citeauthoryear{Robitaille et al.}{2007}]{robitaille07}
	Robitaille, T.~P., Whitney, B.~A., Indebetouw, R. \& Wood, K. 2007,
	ApJS, 169, 328
\bibitem[\protect\citeauthoryear{Romanyuk \& Kudryavtsev}{2008}]{romanyuk08}
	Romanyuk, I.~I. \& Kudryavtsev, D.~O. 2008, Astrophysical Bulletin, 63,
	139
\bibitem[\protect\citeauthoryear{Samus et al.}{2009}]{samus09}
	Samus, N.N., Durlevich, O.V., Kazarovets, E V., Kireeva, N.N., 
	Pastukhova, E.N., Zharova A.V., et al. General Catalog of Variable 
	Stars (GCVS database, Version 2011Jan)
\bibitem[\protect\citeauthoryear{Scargle}{1982}]{scargle82}
	Scargle, J.~D. 1982, ApJ, 263, 835
\bibitem[\protect\citeauthoryear{Schaerer et al.}{1993}]{schaerer}
	Schaerer, D., Charbonnel, C., Meynet, G., Maeder, A. \& Schaller, G.
	1993, A\&AS, 102, 339
\bibitem[\protect\citeauthoryear{Schlegel et al.}{1998}]{schlegel98}
	Schlegel, D.~J., Finkbeiner, D.~P. \& Davis, M. 1998, ApJ, 500, 525
\bibitem[\protect\citeauthoryear{Shulyak et al.}{2004}]{llm}
        Shulyak, D., Tsymbal, V., Ryabchikova, T., St\"utz\, Ch., \& 
	Weiss, W. W. 2004, A\&A, 428, 993
\bibitem[\protect\citeauthoryear{Skiff}{2010}]{skiff2010} 
	Skiff, B. A. 2010, Catalogue of Stellar Spectral Classifications, 
	VizieR On-line Data Catalog: B/mk
\bibitem[\protect\citeauthoryear{Stellingwerf}{1978}]{stellingwerf78} 
	Stellingwerf, R.~F. 1978, ApJ, 224, 953
\bibitem[\protect\citeauthoryear{van Leeuwen}{2007}]{leeuwen} 
	van Leeuwen, F. 2007, A\&A, 474, 653
\bibitem[\protect\citeauthoryear{Watson}{2006}]{watson2006}
	Watson, C.~L. 2006, Society for Astronomical Sciences Annual Symposium,
	25, 47
\bibitem[\protect\citeauthoryear{Wraight et al.}{2011}]{wraight2011}
	Wraight, K.~T., White, G.~J., Bewsher, D. \& Norton, A.~J. 2011, MNRAS,
	in press (arXiv: 1103.0911)
\end{thebibliography}
